\documentclass[iop,apj]{emulateapj}

\begin{document}

\newcommand{\kms}{\ensuremath{\mathrm{km}\,\mathrm{s}^{-1}}}
\newcommand{\MLsun}{\ensuremath{\mathrm{M}_{\sun}/\mathrm{L}_{\sun}}}
\newcommand{\Lsun}{\ensuremath{\mathrm{L}_{\sun}}}
\newcommand{\Msun}{\ensuremath{\mathrm{M}_{\sun}}}
\newcommand{\Aunits}{\ensuremath{\mathrm{M}_{\sun}\,\mathrm{km}^{-4}\,\mathrm{s}^{4}}}
\newcommand{\gevcc}{\ensuremath{\mathrm{GeV}\,\mathrm{cm}^{-3}}}
\newcommand{\etal}{et al.}
\newcommand{\LCDM}{$\Lambda$CDM}
\newcommand{\ML}{\ensuremath{\Upsilon_*}}

\shorttitle{Color--Mass-to-Light Ratio Relations}
\shortauthors{McGaugh \& Schombert}

\title{Color--Mass-to-Light Ratio Relations for Disk Galaxies}

\author{Stacy S. McGaugh}
\affil{Department of Astronomy, Case Western Reserve University, Cleveland, OH 44106}
\email{stacy.mcgaugh@case.edu} 

\and

\author{James M. Schombert}
\affil{Department of Physics, University of Oregon, Eugene, OR 97403}
\email{jschombe@uoregon.edu}

\begin{abstract}
We combine Spitzer $3.6\micron$ observations of a sample of disk galaxies spanning over 10 magnitudes in luminosity
with optical luminosities and colors to test population synthesis prescriptions for computing stellar mass.
Many commonly employed models fail to provide self-consistent results: the stellar mass estimated from the luminosity
in one band can differ grossly from that of another band for the same galaxy.  Independent models agree closely in the 
optical ($V$-band), but diverge at longer wavelengths.  This effect is particularly pronounced in
recent models with substantial contributions from TP-AGB stars. 
We provide revised color--mass-to-light ratio relations that yield self-consistent stellar masses when applied to real galaxies.
The $B-V$ color is a good indicator of the mass-to-light ratio.  
Some additional information is provided by $V-I$, but neither it nor $J-K_s$ are particularly useful for constraining 
the mass-to-light ratio on their own.
In the near-infrared, the mass-to-light ratio depends weakly on color, with typical values of
$0.6\;\MLsun$ in the $K_s$-band and $0.47\;\MLsun$ at $3.6\micron$.
\end{abstract}


\section{Introduction}
\label{sec:intro}

One of the most fundamental properties of a galaxy is its luminosity and the mass of the stars that produce it.
Our understanding of stellar evolution is sufficiently advanced
that it should be possible to compute the luminosity produced by a stellar population ab initio \citep[e.g,][]{BZ03,PEGASE}.
Indeed, there exist in the literature various prescriptions for estimating stellar mass from observed colors
or spectral energy distributions \citep[e.g.,][]{BdJ01,Bell03,Port04,Zib09,IP13}.

A general expectation of population synthesis models is that the relation between mass and light is more nearly constant in the near-infrared (NIR) 
than in the optical part of the spectrum.  This follows from basic considerations:  recent star formation populates the upper main sequence with
luminous, blue stars.  These stars produce copious amounts of optical light from little mass and lead short lives, causing substantial perturbations 
to the average mass-to-light ratio \ML\ of a galaxy.  The degree to which this occurs depends on the IMF and the intensity of a star forming event
relative to the total stellar mass already present.  These effects combine to make the prediction of any particular galaxy's optical mass-to-light ratio
uncertain by a factor of a few.  Young stars contribute rather less in the NIR part of the spectrum, so one expects a closer relation
between light and mass at these wavelengths. Empirically, the scatter in the Tully-Fisher relation declines as one goes from blue to red to
NIR wavelengths \citep{verhTF}, consistent with the expected decrease in scatter in \ML.

In this paper we use $3.6 \micron$ Sptizer Space Telescope photometry \citep{SM2014} 
of a sample of galaxies spanning a large range (ten magnitudes) in luminosity.  
We combine these data with optical colors and luminosities to check the predictions of several population synthesis models.  
Extant models return systematically different stellar masses when applied to the optical and NIR luminosity of the same galaxy.

We describe the data in \S \ref{sec:data}.
In \S \ref{sec:popsynth} we discuss various population synthesis models, and apply them to the data to compute stellar masses in \S \ref{sec:galmass}.
In \S \ref{sec:CMLR} we determine what is required for each model to produce self-consistent results, and present revised color--mass-to-light ratio relations (CMLR)
the provide an improved prescription for estimating stellar mass from photometric data.  We discuss our results and compare them to other constraints
in \S \ref{sec:discuss}, and summarize in \S \ref{sec:conc}. 

\section{Data}
\label{sec:data}

The SINGS \citep{KSINGS,DSINGS} and THINGS \citep{FTHINGS,THINGS} surveys have demonstrated the utility of Sptizer $3.6 \micron$ data for 
constraining the stellar components of star forming galaxies.  Here we wish to sample disk galaxies over as large a range of physical properties as
possible.  To this end, we combine the THINGS data of \citet{THINGS} with new data from from two Spitzer programs.  
One cycle 5 project targeted galaxies to increase the sampling of both higher and lower masses than present in THINGS.  
A cycle 7 snapshot program provides additional photometry for low surface brightness galaxies \citep{SM2014}.  
The combined sample spans ten magnitudes in [3.6] luminosity.  

\begin{deluxetable*}{lccccccr}
\tablewidth{0pt}
\tablecaption{Galaxy Data}
\tablehead{
\colhead{Galaxy} & \colhead{$D$} & \colhead{$M_V$} & \colhead{$M_{[3.6]}$} & \colhead{$B-V$}
& \colhead{$V-I$} & \colhead{$J-K_s$} & \colhead{Refs.}
 }
\startdata
DDO 154  & 4.04	&$-$14.45  &$-$16.41	&0.32   &0.14	&0.34  	&1,2,3,4,5,6 \\
D631-7   & 5.49	&$-$14.50  &$-$16.73	&0.41   &0.55	&\dots 	&1,7\\
D568-2   &21.3	&$-$14.6\phn   &$-$16.81	&0.45   &0.70	&\dots 	&1,3  \\
D572-5   &14.6	&$-$14.56  &$-$16.82	&0.44   &0.52	&\dots 	&1,7 \\
F415-3   &10.4	&$-$15.2\phn   &$-$16.99	&0.62   &0.71	&\dots 	&1,8 \\
DDO 168  & 4.25	&$-$15.70  &$-$17.45	&0.32   &\dots	&\dots 	&1,3,9 \\
F611-1   &25.5	&$-$15.4\phn   &$-$17.92	&0.57   &\dots	&\dots 	&1,8 \\
D500-2   &17.9	&$-$16.38  &$-$18.25	&0.52   &0.42	&\dots 	&1,7 \\
F565-V2  &55.1	&$-$16.2\phn   &$-$18.76	&0.44   &\dots	&\dots 	&1,8 \\
NGC 2366 & 3.27	&$-$16.82  &$-$18.90	&0.54   &0.52	&0.84 	&2,5,6,9 \\
D723-5   &27.7	&$-$16.9\phn   &$-$19.21	&0.55   &0.75	&\dots 	&1,3 \\
F563-V1  &57.6	&$-$17.2\phn   &$-$19.79	&0.23   &0.83	&\dots 	&1,8 \\
IC 2574  & 3.91	&$-$17.70  &$-$20.15	&0.42   &0.67	&0.58 	&2,5,6,8 \\
F563-1   &52.2	&$-$17.8\phn   &$-$20.40	&0.40   &0.86	&\dots 	&1,8 \\
F574-2   &92.3	&$-$18.3\phn   &$-$20.50	&0.58   &\dots	&\dots 	&1,8 \\
NGC 2976 & 3.58	&$-$17.80  &$-$20.52	&0.55   &0.67	&0.79 	&2,5,6,10 \\
F568-V1  &84.8	&$-$18.6\phn   &$-$20.82	&0.47   &0.70	&\dots 	&1,8,11 \\
F561-1   &69.8	&$-$18.3\phn   &$-$20.88	&0.69   &0.72	&\dots 	&1,3,8 \\
F577-V1  &113.	&$-$18.7\phn   &$-$20.95	&0.50   &1.07	&\dots 	&1,3,8 \\
NGC 1003 & 10.2 &$-$18.79  &$-$21.12	&0.42   &\dots	&0.73 	&2,5,6 \\
UGC 5005 & 57.1	&$-$18.7\phn   &$-$21.18	&0.35   &\dots	&\dots 	&1,8 \\
F574-1   &100.	&$-$19.1\phn   &$-$21.37	&0.51   &\dots	&\dots 	&1,8 \\
F568-1   & 95.5	&$-$18.9\phn   &$-$21.38	&0.52   &0.70	&\dots 	&1,8,11 \\
NGC 7793 & 3.61	&$-$18.86  &$-$21.46	&0.63   &0.20	&0.68 	&2,5,6 \\
UGC 128  &58.5	&$-$19.3\phn   &$-$21.88	&0.63   &0.68	&\dots 	&1,11 \\
NGC 2403 & 3.16	&$-$19.14  &$-$21.97	&0.39   &1.03	&0.75 	&2,5,6,10 \\
NGC 925  &9.43	&$-$19.97  &$-$22.30	&0.50   &0.75	&0.83 	&2,5,6,10 \\
NGC 2903 & 8.9	&$-$20.82  &$-$22.74	&0.55   &1.16	&0.91 	&2,5,6,12 \\
NGC 3198 & 13.8	&$-$20.40  &$-$23.00	&0.43   &1.02	&0.92 	&2,5,6,12 \\
NGC 3621 & 6.56	&$-$19.74  &$-$23.04	&0.52   &0.81	&0.83 	&2,5,6 \\
NGC 3521 & 8.0	&$-$20.65  &$-$24.19	&0.68   &1.18	&0.93 	&2,5,6,10 \\
NGC 3031 & 3.65	&$-$21.08  &$-$24.28	&0.82   &1.31	&0.88 	&2,5,6,10 \\
NGC 5055 & 8.99	&$-$21.22  &$-$24.60	&0.64   &1.21	&0.95 	&2,5,6,10 \\
NGC 2998 & 68.3	&$-$22.36  &$-$24.77	&0.45   &\dots	&0.99 	&2,5,6 \\
NGC 2841 & 14.1	&$-$21.57  &$-$24.88	&0.74   &1.35	&0.93 	&2,5,6,10 \\
NGC 6674 & 51.9	&$-$22.17  &$-$25.17	&0.57   &\dots	&0.86 	&2,5,6 \\
NGC 7331 & 14.9	&$-$21.63  &$-$25.30	&0.63   &1.36	&1.00 	&2,5,6,10 \\
NGC 801  & 75.3 &$-$22.30  &$-$25.33	&0.61   &\dots	&1.05 	&2,5,6,9 \\
NGC 5533 & 59.4	&$-$22.16  &$-$25.47	&0.77   &\dots	&0.94 	&2,5,6 \\
UGC 2885 & 75.9	&$-$23.30  &$-$25.94	&0.47   &\dots	&0.88 	&1,5,6
\enddata
\label{galdata}
\tablerefs{Spitzer [3.6] magnitudes: 1.~\citet{SM2014} 2.~\citet{THINGS}.
Ancillary data: 3.~\citet{SMM11} 4.~\citet{Dale07}
5.~\citet{RC3} 6.~\citet{2MASS} 7.~\citet{trach} 8.~\citet{MB94} 9.~\citet{Mak99}
10.~\citet{MM09} 11.~\citet{dBHB95} 12.~\citet{FD08}.
\tablecomments{Galaxy photometric data in order of increasing [3.6] luminosity.
Adopted distances are in Mpc.  $H_0 = 75\;\kms\,\mathrm{Mpc}^{-1}$ is assumed when
no direct determination is available.  We adopt $M_{\sun}^V = 4.83$, $M_{\sun}^I = 4.08$, and $M_{\sun}^{[3.6]} = 3.24$.}
}
\end{deluxetable*}

The [3.6] luminosities of galaxies in the THINGS sample have been adopted from the mass models of \citet{THINGS}.
Only the total luminosity is used here.  No distinction is made between bulge and disk components.  

\begin{deluxetable*}{llccccccccccc}
\tablewidth{0pt}
\tablecaption{Population Synthesis CMLR}
\tablehead{
\colhead{Model} & \colhead{IMF} & \colhead{$a_V$} & \colhead{$b_V$}  & \colhead{$a_{I}$}  & \colhead{$b_{I}$} & \colhead{$a_{K}$}  & \colhead{$b_{K}$} & 
\colhead{$\Upsilon_{0.6}^V$} & \colhead{$\Upsilon_{0.6}^I$} & \colhead{$\Upsilon_{0.6}^K$} & \colhead{$\Upsilon_{0.6}^{[3.6]}$} & \colhead{AGB}
}
\startdata
Bell \etal\ (2003)  & Scaled Salpeter & $-$0.628 & 1.305 & $-$0.399 & 0.824 & $-$0.206 & 0.135 & 1.43 &1.25 & 0.73 & 0.62 & old \\
Portinari \etal\ (2004) & Kroupa (1998) &$-$0.654 & 1.290 & $-$0.537 & 0.970 & $-$0.736 & 0.730 & 1.32 &1.11 & 0.50 & 0.41 & old \\
Zibetti \etal\ (2009)  & Chabrier (2003) & $-$1.075 & 1.837 & $-$1.003 & 1.475 & $-$1.390 & 1.176 & 1.07 &0.76 & 0.21 & 0.14 & new \\
Into \& Portinari (2013) & Kroupa (1998) & $-$0.900 & 1.627 & $-$0.782 & 1.294 & $-$1.020 &1.054 & 1.19 &0.99 & 0.41 & 0.33 & new 
\enddata
\tablecomments{Stellar mass-to-light ratios in the $V$, $I$, and $K$-bands as given by various population synthesis models
in solar units through the formula $\log \ML^i = a_i + b_i (B-V)$.  For reference, the mass-to-light ratios
predicted by each model for $B-V = 0.6$ are also given.  The AGB column denotes whether the model includes older or newer
\citep{Marigo2008} prescriptions for TP-AGB stars.}
\label{popsynthML} 
\end{deluxetable*}

The remainder of the assembled sample is composed of new Spitzer observations obtained by ourselves \citep{SM2014}.
The Spitzer data have been analyzed with the ARCHANGEL surface photometry package \citep{ARCHANGEL}.  Elliptical isophotes have been fit to
the data and integrated magnitudes determined from asymptotic fits to curves of growth.  Special care has been taken to exclude 
foreground stars and background galaxies and replace the masked region with an estimate of the galaxy light based on surrounding pixels.  
IRAC is a sensitive instrument, and many background galaxies shine through the disks of the target galaxies at $3.6\micron$.
Careful cleaning of these contaminants is essential to accurate photometry.  
Once this step is taken, total magnitudes can be determined to a few hundredths of a magnitude.  
Colors and magnitudes are corrected for Galactic extinction using the calibration of \citet{SF2011}.
Internal extinction corrections follow the RC3 convention \citep{RC3}, 
but this is only substantial ($\approx 0.2$ mag.) in the brightest few galaxies.

We assume that the observed NIR light is stellar in origin, and make no attempt to correct for non-stellar contamination (e.g., PAH emission).  
This is small at $3.6\micron$ \citep{JHKim2012,Meidt2012}.  Indeed, for the galaxies detected by IRAS, the observations of \citet{JHKim2012}
can be used to estimate the amount of contamination expected from the $3.3\micron$ PAH feature.  In all cases, it is expected to be $< 2\%$ of the
observed flux, and usually much less than 1\% (J.H.~Kim, private communication).  It will be less for the low surface brightness galaxies not detected by IRAS.

The data are presented in Table~\ref{galdata}.  For each galaxy, we assign a distance and compute the corresponding absolute
magnitude in the optical $V$-band and NIR Sptizer IRAC [3.6] band.  Distances are taken from the direct measurements tabulated
in the Extragalactic Distance Database \citep{EDD} when available.  When no direct distance measurement is known, a Hubble flow distance
assuming $H_0 = 75\;\kms\,\mathrm{Mpc}^{-1}$ is adopted.  Measured colors are given where known.  The majority of the
sample have observed $B-V$ colors. For brighter galaxies, $J-K_s$ can readily be extracted from 2MASS \citep{2MASS}.  The lower
surface brightness galaxies are typically not detected\footnote{DDO 154 is only marginally detected by 2MASS, so we are skeptical
of its $J-K_s$ color.  Colors measured by ourselves are good to a few hundredths of a magnitude, and inter-comparison of other
modern photometry \citep[e.g.][]{Dale07,MM09} is similarly encouraging.} by 2MASS.  
For many, $V-I$ has been observed \citep{SMM11}.  The colors quoted there were based on fixed apertures; here we adopted
a weighted average to better represent the extended low surface brightness portions of the disks.  
The isophotal weighted and aperture colors rarely differed by more than 0.05 mags except where a prominent bulge component was present.
References to the sources of these data are given in the final column of Table~\ref{galdata}. 

\section{Population Synthesis Models}
\label{sec:popsynth}

Our knowledge of stellar evolution is sufficiently advanced to enable the ab initio calculation of the spectral energy distribution (SED) of stellar populations.
Considerable effort has gone into the development of stellar population synthesis models that do just this \citep[e.g.,][]{BZ03}.
Outstanding progress has been made, and it has become standard practice to quote stellar masses for galaxies based on fits to
multi-color data with such models.

\begin{figure*}
\epsscale{1.0}
\plotone{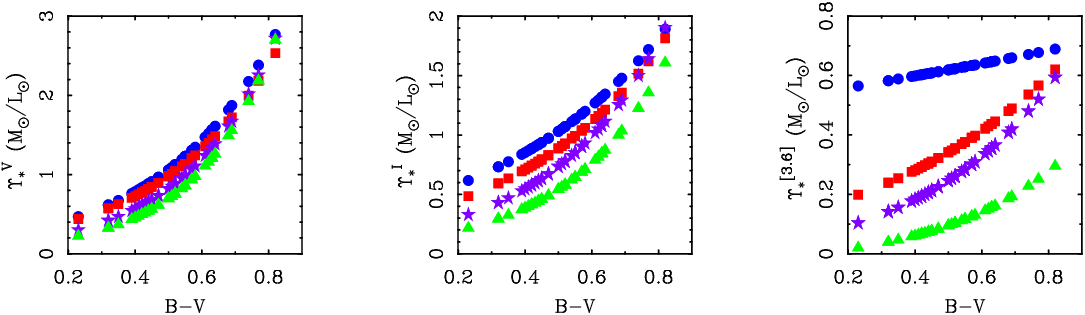}
\caption{The relation between $B-V$ color and the stellar mass-to-light ratio in the $V$-band (left), $I$-band (center) 
and the Spitzer [3.6] band (right) from the population synthesis models of \citet{Bell03} with their scaled Salpeter IMF (circles),
\citet{Port04} with a \citet{kroupa} IMF (squares), \citet{Zib09} with a \citet{chabrier} IMF (triangles), and \citet{IP13} 
with a \citet{kroupa} IMF (stars).  The formula of \citet[equation \ref{eqn:Oh}]{OHTHINGS} is used to convert between $K_s$ and [3.6].
Note the large disparity between models in the NIR.  
\label{MLcolor}}
\end{figure*}

The accuracy with which stellar masses can be estimated is debatable, but is probably no better than a factor of two. 
Outstanding problems include uncertainty in the IMF, variations in the star formation histories of galaxies, 
the distribution of stellar metallicities, and the contribution of  stars in bright but short-lived phases of evolution (e.g., TP-AGB stars).
Here we compare the predictions of various models with each other.  We also check their internal self-consistency to gage the extent to which
the same model predicts the same stellar mass for the same galaxy when the luminosity is measured in different bands.

A simple approach to estimating the stellar mass of a galaxy is to assume a single, constant mass-to-light ratio \ML\ such that $M_* = \ML L$.
This is a crude approximation, as we expect the mass-to-light ratio of a population to vary with age and, to a lesser extent, with metallicity. 
For example, using a mutli-metallicity model \citep{rakosschombert09a,SM2014b}, we find that a 12 Gyr old stellar population of solar metallicity 
has $\ML^V = 2.8\;\MLsun$, while a stellar population with the same age but peak [Fe/H] $= -1.5$ has $\ML^V = 1.8\;\MLsun$.  
That same solar metallicity stellar population has $\ML^V$ of only $0.4\;\MLsun$ at an age of 1 Gyr.  

A more sophisticated approach is to use a population synthesis model to construct a CMLR.
This relates the mean mass-to-light ratio $\ML^i$ in band $i$ to a color $(m_j - m_k)$ through
\begin{equation}
\log \ML^i = a_i + b_i (m_j - m_k).
\label{eqn:popsynth}
\end{equation}
The bands $i, j, k$ can be independent, but need not be.  That is, sometimes band $i = j$ or $k$.

Using a color as a mass-to-light ratio estimator reduces to the simple approach if the slope $b$ is small for some band $i$. 
Variation of \ML\ with color is expected to be minimized in the NIR.  Similarly, we expect optical colors to provide an indicator of \ML.
For the multi-metallicity models of \citet{rakosschombert09a}, we find that the solar metallicity model changes in color as it ages from 1 to 12 Gyr by
$\Delta(B-V) = 0.37$ and $\Delta(J-K) = 0.03$.  Thus we expect $B-V$ to be a more sensitive indicator of \ML\ than $J-K_s$. 

There should be some intrinsic scatter about the mean CMLR. 
This scatter ultimately limits the accuracy achievable by this approach, but  
is expected to be minimized in the NIR \citep{BdJ01}.
One might hope to do better by using multi-color information \citep[e.g.,][]{Zib09}, or fitting the entire SED.   
The accuracy of stellar masses inferred from SED fitting is however limited by the fidelity 
of the population synthesis model to which the SED is fit.  This approach will suffer systematic error if a model differs from reality 
as a function of wavelength.  

The coefficients $a_i$ and $b_i$ are given in Table~\ref{popsynthML} for several representative models for the $i=V$, $I$, and $K$ bands with 
$B-V$ color.  These particular choices are made because the most data are available in these bands.  As we shall see, $B-V$ has some value 
as a predictor of \ML, while $V-I$ and $J-K_s$ do not.

The models typically stop at $K$ while we now have a good deal of Spitzer [3.6] photometry.  
We relate the population synthesis predicted $K_s$-band mass-to-light ratio
$\ML^K$ to $\ML^{[3.6]}$ using the relation of \citet{OHTHINGS}:
\begin{equation}
\ML^{[3.6]} = 0.92 \ML^K - 0.05.
\label{eqn:Oh}
\end{equation}
This relation is obtained from population synthesis models, in the same spirit as the CMLR in Table \ref{popsynthML}.

To convert between $\ML^K$ to $\ML^{[3.6]}$ in the data, we assume $\ML^K = 1.29 \ML^{[3.6]}$.
This follows from the mean color $K_s-[3.6] = 0.31 \pm 0.11$ that we obtain for the 74 galaxies of \citet{DSINGS}.
In the larger S4G sample, we find a weak anticorrelation between $K_s-[3.6]$ and $B-V$ \citep{SM2014b} such that 
the conversion factor would vary between 1.39 for $B-V = 0.3$ and 1.18 for $B-V = 0.8$.
This range of variation is within the scatter of $K_s-[3.6]$ at a given $B-V$, so we only employ the mean 
value and do not attempt to estimate $K_s-[3.6]$ from this weak correlation with $B-V$.  

Note that the conversion for the data is not identical to that for the models.  
As we will see, the models do not perform well in reproducing the data in the NIR.  
We therefore choose to keep the two separate,
making the model conversion with a model result and the data conversion with the mean of the data.

\begin{figure*}
\epsscale{1.0}
\plotone{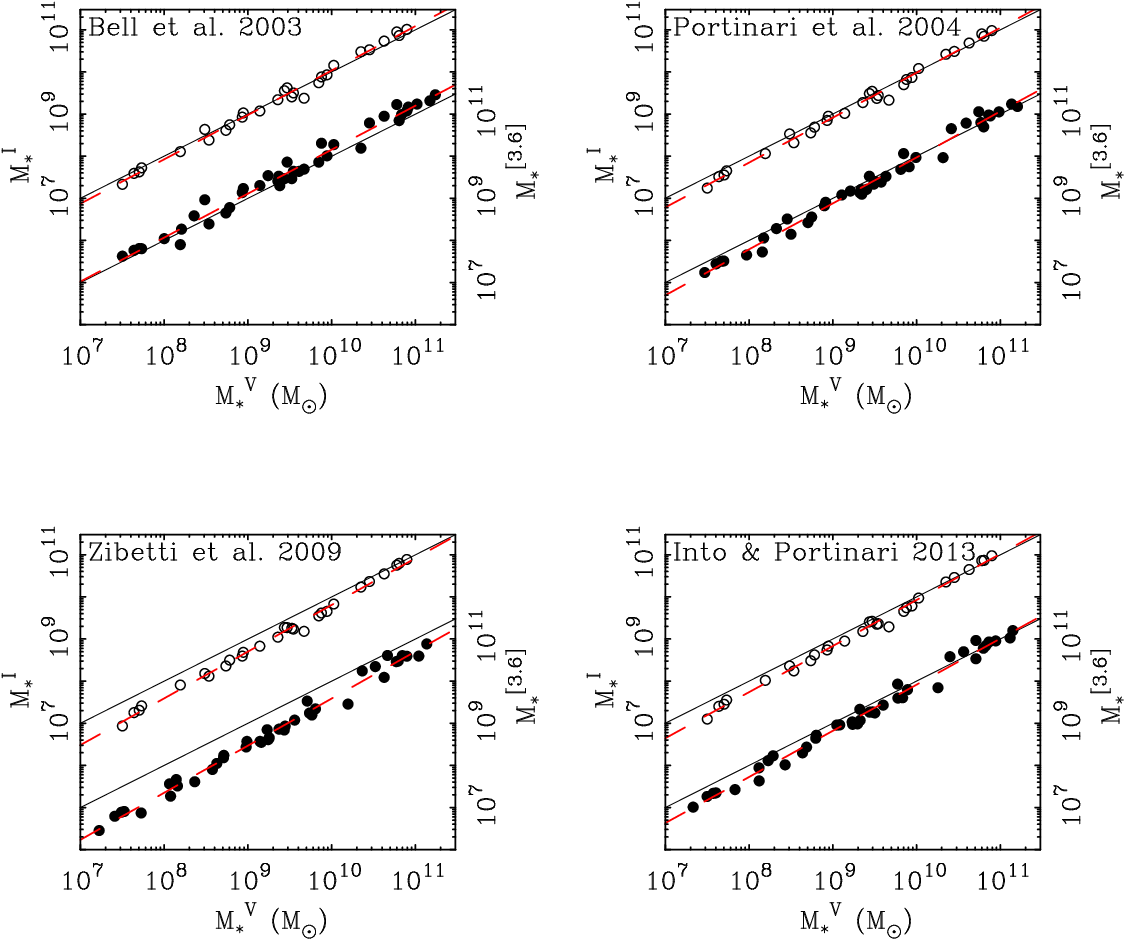}
\caption{Stellar masses (Table~\ref{popsynth}) estimated by population synthesis models \citep[Table~\ref{popsynthML}]{Bell03,Port04,Zib09,IP13}.  
For each model, the mass estimated from
the either the $I$-band (open circles) or [3.6] luminosity (filled circles) of each galaxy is plotted against that estimated from the $V$-band luminosity.  
The two cases are offset for clarity.  If the models were perfect 
the data would follow the solid lines of unity, modulo the expected intrinsic scatter in the relation between the mass-to-light ratio and color.  
Dashed lines show fits to the data (Table \ref{MtoMlines}) quantifying the deviation from this ideal.
These all have slopes greater than unity, indicating that the sensitivity of the mass-to-light ratio to color in $I$ and [3.6] is overstated
relative to that in $V$.  The models also tend to over-predict the [3.6] luminosity relative to the optical luminosity, with the 
exception of the model of \citet{Bell03}, which underestimates it.}
\label{MvM}
\end{figure*}

Fig.~\ref{MLcolor} shows the mass-to-light ratios for the models given in Table~\ref{popsynthML}.
Rather than simply show a line for each model, we plot the galaxy data to emphasize the beads-on-a-string nature of this approach
to estimating stellar mass:  a single color may provide a reasonable estimate of the mean mass-to-light ratio, but it cannot reproduce the
intrinsic scatter that one expects from variations in star formation histories.  
The scatter is expected to be from 0.1 dex \citep{BdJ01} to 0.15 dex \citep{Port04} in the $K$-band, and larger in the optical bands.
This is an over-simplification, as the scatter may be a function of color, with larger scatter likely in very blue, actively star forming
systems.  Additionally, the mean CMLR can 
bend \citep[the slope becomes much steeper for $B-V < 0.55$ in the models of][]{Port04} or even 
bifurcate: the line of \citet{Bell03} splits the difference between distinct branches of high and low $\ML^K$ at blue colors in their Fig.~20.

Models will differ if they adopt different evolutionary tracks or a different IMF.  Both matter for the models considered here.  
As pointed out by \citet{BdJ01}, changes in the IMF to include more or fewer low mass stars serve mostly to change the mass without 
much altering the luminosity or color, so to a decent approximation can be treated as multiplicative shifts in \ML.
For specificity, we adopt the scaled Salpeter IMF of \citet{BdJ01} for the model of \citet{Bell03}.  
We adopt the Kroupa IMF \citep{kroupa} for the model of \citet{Port04} and \citet{IP13}, while \citet{Zib09} uses the Chabrier IMF \citep{chabrier}.

The models all give a similar run of \ML\ with color in the optical (left panel of Fig.~\ref{MLcolor}), 
with small offsets\footnote{The models of \citet{BdJ01} with a scaled Salpeter IMF are barely distinguishable from those
of \citet{Port04} with a Kroupa IMF, so we consider only the latter.}
 for the different IMFs as well as other detailed differences.
The agreement degrades as we move to redder wavelengths (middle and right panels of Fig.~\ref{MLcolor}). 
There is a huge disparity in the NIR.  The model of \citet{Bell03} has a relatively flat slope  
with a high normalization $\ML^{[3.6]} > 0.55\;\MLsun$, 
while that of \citet{Zib09} has a steep dependence on color (even in the NIR) and a low 
normalization $\ML^{[3.6]} < 0.3\;\MLsun$ for the reddest galaxies and $< 0.1\;\MLsun$ for many blue galaxies.
The model of \citet{IP13} is intermediate between that of \citet{Zib09} and \citet{Port04}.

The chief difference in evolutionary tracks between the various models considered here is the inclusion of 
a large contribution from TP-AGB stars by \citet{Zib09} and \citet{IP13} as advocated by \citet{maraston05}.  
These stars are in the latest stages of evolution, being short-lived and rare, but quite bright.  Their contribution to the integrated
luminosity of stellar populations is most pronounced in the NIR, where they greatly enhance the predicted luminosity of galaxies while doing little
to alter the predictions of previous generations of models in the optical portion of the spectrum.
This results in the low mass-to-light ratios of the models\footnote{\citet{Zib09} advocate using multiband colors to estimate mass-to-light ratios,
and their single-color coefficients that are reproduced in Table~\ref{popsynthML} are only approximations made to facilitate the sort of comparison made here.}
of \citet{Zib09} and \citet{IP13} in the right hand panel of Fig.~\ref{MLcolor}.
These models will obviously give rather different estimates of the stellar mass, especially when applied in the NIR.

\section{Galaxy Stellar Masses}
\label{sec:galmass}

We use the models in Table~\ref{popsynthML} with the data in Table~\ref{galdata} to compute the stellar masses of sample galaxies.  
These are reported in Table~\ref{popsynth}.  For each galaxy, we use the observed $B-V$ color to predict the mass-to-light ratio separately in 
$V$, $I$, and [3.6] for each model.  We then use the corresponding luminosity to obtain a stellar mass estimate. This results in twelve 
distinct mass estimates for each galaxy: three for each of the four models.  
A similar exercise could be performed using other colors, but we find $V-I$ and $J-K$ to be less satisfactory\footnote{Attempts
to build the equivalent of Table \ref{popsynth} with these colors not only limit the dataset since there are fewer measurements, 
but also produce noisy and sometimes unphysical results.} than $B-V$ as primary \ML\ estimators.  We will consider their use as a second color term later.

\begin{deluxetable*}{lcccccccccccc}
\tablewidth{0pt}
\tablecaption{Stellar Masses from Population Synthesis Models}
\tablehead{
& \multicolumn{3}{c}{B03}
& \multicolumn{3}{c}{P04} & \multicolumn{3}{c}{Z09} & \multicolumn{3}{c}{IP13} \\
\colhead{Galaxy}
& \colhead{M$_*^V$} & \colhead{M$_*^I$} & \colhead{M$_*^{[3.6]}$}
& \colhead{M$_*^V$} & \colhead{M$_*^I$} & \colhead{M$_*^{[3.6]}$}
& \colhead{M$_*^V$} & \colhead{M$_*^I$} & \colhead{M$_*^{[3.6]}$}
& \colhead{M$_*^V$} & \colhead{M$_*^I$} & \colhead{M$_*^{[3.6]}$}
}
\startdata
DDO 154         &7.50   &7.33   &7.63   &7.47   &7.24   &7.24   &7.22   &6.94   &6.45   &7.33   &7.10   &7.01 \\
D631-7          &7.64   &7.59   &7.77   &7.67   &7.51   &7.45   &7.41   &7.25   &6.79   &7.50   &7.40   &7.26 \\
D568-2          &7.73   &7.73   &7.80   &7.70   &7.65   &7.51   &7.52   &7.41   &6.91   &7.60   &7.55   &7.35 \\
D572-5          &7.70   &7.63   &7.81   &7.67   &7.55   &7.51   &7.49   &7.31   &6.89   &7.57   &7.45   &7.34 \\
F415-3          &8.19   &8.11   &7.90   &8.16   &8.06   &7.72   &8.08   &7.91   &7.27   &8.12   &8.02   &7.63 \\
DDO 168         &8.00   &\dots  &8.04   &7.97   &\dots  &7.65   &7.72   &\dots  &6.87   &7.83   &\dots  &7.43 \\
F611-1          &8.21   &\dots  &8.27   &8.17   &\dots  &8.06   &8.06   &\dots  &7.56   &8.12   &\dots  &7.94 \\
D500-2          &8.53   &8.38   &8.39   &8.50   &8.32   &8.15   &8.36   &8.12   &7.61   &8.43   &8.24   &8.01 \\
F565-V2         &8.36   &\dots  &8.58   &8.36   &\dots  &8.28   &8.15   &\dots  &7.67   &8.23   &\dots  &8.11 \\
NGC 2366        &8.74   &8.61   &8.65   &8.70   &8.55   &8.42   &8.58   &8.36   &7.90   &8.64   &8.48   &8.30 \\
D723-5          &8.78   &8.75   &8.78   &8.75   &8.69   &8.56   &8.63   &8.50   &8.05   &8.69   &8.62   &8.43 \\
F563-V1         &8.48   &8.63   &8.96   &8.45   &8.53   &8.51   &8.16   &8.18   &7.51   &8.29   &8.36   &8.23 \\
IC 2574         &8.93   &8.93   &9.14   &8.90   &8.85   &8.82   &8.71   &8.60   &8.18   &8.80   &8.74   &8.64 \\
F563-1          &8.95   &9.03   &9.23   &8.91   &8.95   &8.90   &8.71   &8.68   &8.24   &8.80   &8.83   &8.72 \\
F574-2          &9.38   &\dots  &9.30   &9.35   &\dots  &9.10   &9.24   &\dots  &8.61   &9.30   &\dots  &8.99 \\
NGC 2976        &9.14   &9.07   &9.30   &9.11   &9.02   &9.08   &8.99   &8.83   &8.57   &9.05   &8.95   &8.96 \\
F568-V1         &9.36   &9.34   &9.41   &9.32   &9.27   &9.13   &9.16   &9.04   &8.55   &9.24   &9.18   &8.98 \\
F561-1          &9.52   &9.41   &9.47   &9.49   &9.37   &9.34   &9.44   &9.25   &8.93   &9.47   &9.35   &9.27 \\
F577-V1         &9.44   &9.55   &9.47   &9.40   &9.49   &9.21   &9.26   &9.27   &8.65   &9.33   &9.41   &9.07 \\
NGC 1003        &9.37   &\dots  &9.52   &9.34   &\dots  &9.21   &9.14   &\dots  &8.57   &9.23   &\dots  &9.03 \\
UGC 5005        &9.24   &\dots  &9.54   &9.21   &\dots  &9.17   &8.98   &\dots  &8.44   &9.08   &\dots  &8.96 \\
F574-1          &9.61   &\dots  &9.64   &9.58   &\dots  &9.39   &9.43   &\dots  &8.84   &9.50   &\dots  &9.25 \\
F568-1          &9.54   &9.50   &9.64   &9.51   &9.44   &9.40   &9.37   &9.24   &8.86   &9.44   &9.36   &9.26 \\
NGC 7793        &9.67   &9.38   &9.69   &9.63   &9.33   &9.52   &9.56   &9.18   &9.07   &9.60   &9.29   &9.43 \\
UGC 128         &9.85   &9.74   &9.86   &9.81   &9.70   &9.69   &9.73   &9.55   &9.24   &9.78   &9.66   &9.60 \\
NGC 2403        &9.47   &9.62   &9.86   &9.44   &9.54   &9.52   &9.23   &9.27   &8.85   &9.32   &9.42   &9.33 \\
NGC 925         &9.95   &9.93   &10.01  &9.91   &9.87   &9.75   &9.76   &9.65   &9.19   &9.83   &9.79   &9.61 \\
NGC 2903        &10.35  &10.48  &10.19  &10.32  &10.42  &9.97   &10.20  &10.23  &9.46   &10.25  &10.35  &9.85 \\
NGC 3198        &10.03  &10.16  &10.28  &9.99   &10.08  &9.97   &9.81   &9.83   &9.34   &9.89   &9.974  &9.80 \\
NGC 3621        &9.88   &9.88   &10.31  &9.84   &9.819  &10.06  &9.71   &9.62   &9.53   &9.77   &9.743  &9.93 \\
NGC 3521        &10.45  &10.53  &10.79  &10.42  &10.49  &10.65  &10.37  &10.36  &10.24  &10.40  &10.46  &10.58 \\
NGC 3031        &10.81  &10.86  &10.85  &10.77  &10.85  &10.80  &10.80  &10.79  &10.48  &10.80  &10.87  &10.78 \\
NGC 5055        &10.63  &10.73  &10.95  &10.59  &10.69  &10.78  &10.52  &10.55  &10.35  &10.56  &10.65  &10.70 \\
NGC 2998        &10.84  &\dots  &10.99  &10.80  &\dots  &10.70  &10.63  &\dots  &10.09  &10.71  &\dots  &10.53 \\
NGC 2841        &10.90  &11.01  &11.07  &10.86  &10.98  &10.98  &10.84  &10.89  &10.61  &10.86  &10.98  &10.93 \\
NGC 6674        &10.92  &\dots  &11.17  &10.88  &\dots  &10.96  &10.77  &\dots  &10.46  &10.83  &\dots  &10.84 \\
NGC 7331        &10.78  &10.95  &11.23  &10.74  &10.90  &11.06  &10.67  &10.75  &10.61  &10.71  &10.86  &10.97 \\
NGC 801         &11.02  &\dots  &11.24  &10.98  &\dots  &11.05  &10.90  &\dots  &10.59  &10.94  &\dots  &10.95 \\
NGC 5533        &11.17  &\dots  &11.31  &11.14  &\dots  &11.24  &11.14  &\dots  &10.88  &11.15  &\dots  &11.20 \\
UGC 2885        &11.24  &\dots  &11.46  &11.20  &\dots  &11.18  &11.04  &\dots  &10.60  &11.12  &\dots  &11.02
\enddata
\tablecomments{Masses are base ten logarithms in \Msun.  For each population model, the mass estimate
from the $V$-band luminosity is given first, then that from the $I$-band, then the [3.6] luminosity.
The models used are those of
\citet[B03]{Bell03}, \citet[P04]{Port04}, \citet[Z09]{Zib09}, and \citet[IP13]{IP13}.
}
\label{popsynth}
\end{deluxetable*}

We can now compare stellar mass estimate from different population synthesis models.  
The external consistency of the models is fairly good in the optical:  examination of Table~\ref{popsynth} shows that $M_*^V$ is usually similar across the
board.  There are small offsets owing largely to differences in the adopted IMF, which mostly affect the normalization $a_V$.  
There are also small differences stemming from differences in the slope $b_V$.  More recent models predict a somewhat steeper slope with color.
These small differences are to be expected.  All in all, the external consistency between models in the $V$-band is encouraging.

We can also check each model for internal self-consistency from band to band.
If all is well, the stellar mass estimated for the same galaxy will be the same irrespective of whether the luminosity is measured in the optical or NIR.
Intrinsic scatter in the CMLR precludes this from ever being exactly true, but the sample is large enough that we can check whether it is true 
on average.  This is done in Fig.~\ref{MvM}, which shows the stellar mass estimated from the $I$-band and [3.6] luminosity plotted against that estimated
from the $V$-band luminosity.

The approximate consistency between models in $V$ does not hold in the NIR.  
Indeed, most of the models are not internally self-consistent:  for a given stellar mass, most models over-predict 
the infrared luminosity relative to the optical luminosity.  In other words, the predicted infrared mass-to-light ratios are too small relative to those
in the optical.

\begin{deluxetable}{lcccc}
\tablewidth{0pt}
\tablecaption{Self-Consistent Stellar Masses}
\tablehead{
\colhead{Model} & \colhead{$\log(\mathrm{M}_{0}^I)$}  & \colhead{$B_{I}$} & \colhead{$\log(\mathrm{M}_{0}^{[3.6]})$}  & \colhead{$B_{[3.6]}$} 
}
\startdata
Bell \etal\ (2003)  & 9.393 & 1.054 & 6.431 & 1.043  \\
Portinari \etal\ (2004) & 10.262 & 1.067 & 10.315 & 1.091 \\
Zibetti \etal\ (2009)  & 11.719 & 1.109 & 13.531 & 1.118  \\
Into \& Portinari (2013) & 10.779 & 1.094 & 10.913 & 1.095 
\enddata
\tablecomments{Fits to the data in Fig.\ \ref{MvM} that reconcile stellar masses from the $I$-band and [3.6] with those from the $V$-band
such that $\log(\mathrm{M}_*^j/\mathrm{M}_0) = B_j \log(\mathrm{M}_*^V/\mathrm{M}_0)$, where $B_j$ is the slope of the fitted line, 
and M$_0$ the mass where the $V$-band and band $j = I$ or [3.6] intersect.}
\label{MtoMlines} 
\end{deluxetable}

The data cover many decades in luminosity.  Consequently, even a small offset from the line of equality in the logarithmic Fig.~\ref{MvM} 
corresponds to a serious misestimation of the mass-to-light ratio.  The semi-empirical model of \citet{Bell03} is most nearly self-consistent.
Indeed, it provides a very good match between $V$ and $I$ bands, with only a small tendency 
to over-predict the stellar mass from the NIR.  The other models all under-predict
the stellar mass from the NIR luminosity relative to that in the $V$-band.  This problem is particularly severe in the model of \citet{Zib09},
which also suffers the same problem in the $I$-band.  The offset and change in slope seen in Fig.~\ref{MvM} is an indication that both the
intercept $a_i$ and slope $b_i$ are misestimated.

\begin{figure*}
\epsscale{1.0}
\plotone{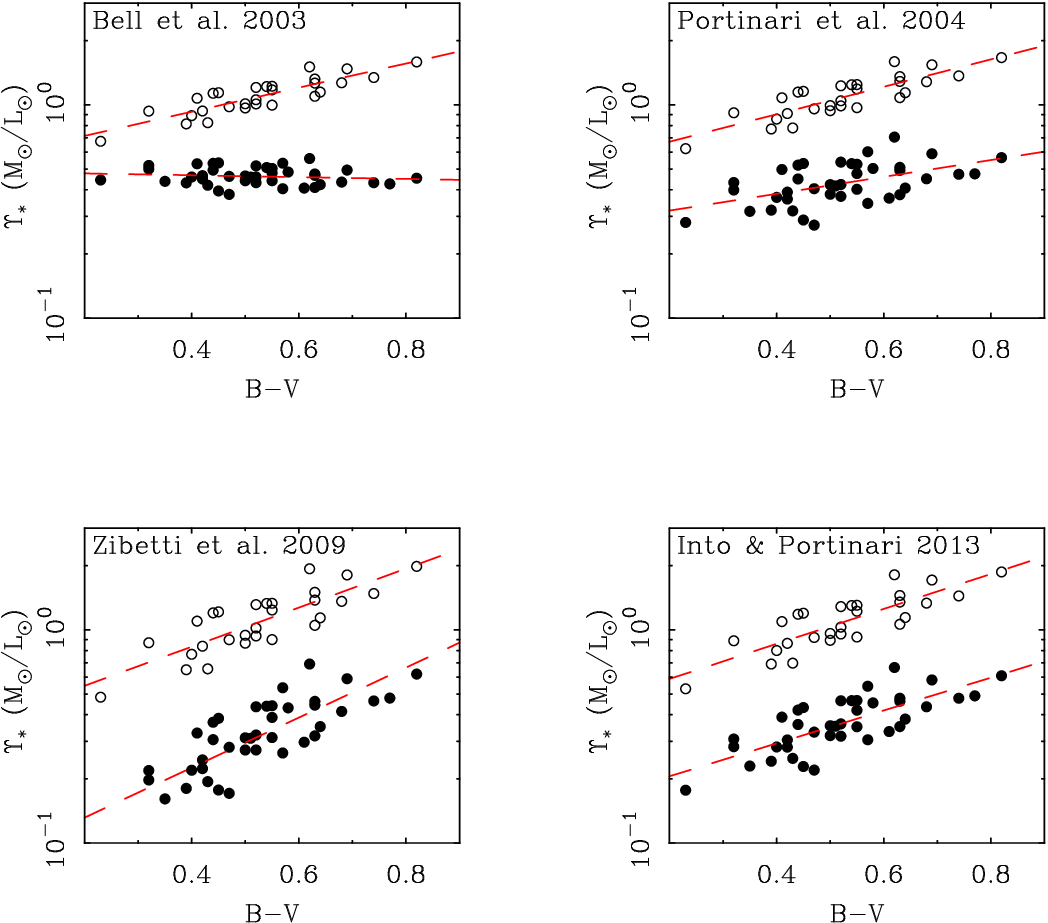}
\caption{Stellar mass-to-light ratios in the $I$-band (open circles) and [3.6] (filled circles) as a function of $B-V$ color.
The models of \citet{Bell03,Port04,Zib09,IP13} have been revised according to the fits in Table \ref{MtoMlines}.
Fits to the data (dashed lines) are given in Table \ref{popsynthrevised}.  There is scatter in the data because
the $V-I$ and $V-[3.6]$ colors of a galaxy vary at a given $B-V$.  Note that the bluest galaxies falls off the graph
with $\ML^{[3.6]} < 0.1\;\MLsun$ in the lower left plot.
\label{MLpop_VIK}}
\end{figure*}

The chief difference between older and more recent models is the prescription for TP-AGB stars \citep{Marigo2008}.
This prescription appears to grossly overstate the contribution of TP-AGB stars to the NIR luminosity of real galaxies.
Apparently the contribution of these luminous but short-lived stars to the integrated energy budget has been overestimated.  

\begin{figure*}
\epsscale{1.0}
\plotone{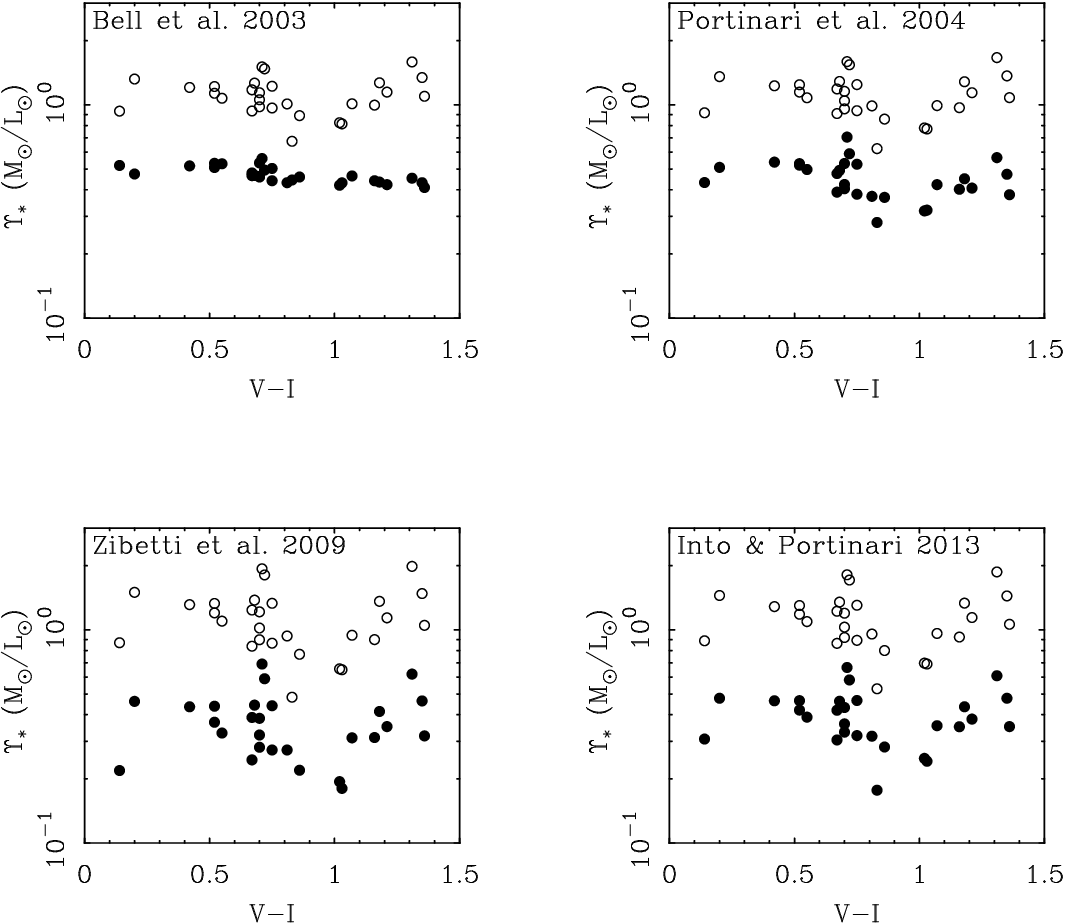}
\caption{Stellar mass-to-light ratios in the $I$-band (open circles) and [3.6] (filled circles) as a function of $V-I$ color.
The models of \citet{Bell03,Port04,Zib09,IP13} have been revised according to the fits in Table \ref{MtoMlines}.
Contrary to the case of $B-V$ (Fig.~\ref{MLpop_VIK}), there is little correlation:  $V-I$ is not a good primary indicator of \ML,
though it does have some value as a secondary indicator when combined with $B-V$ (Fig.~\ref{MLpoptweak_VIK}).
\label{MLpop_VIK_VI}}
\end{figure*}

Our conclusion concerning the ratio of NIR to optical luminosity is consistent with the findings of other workers.
\citet{Melbourne12} reach a similar conclusion from resolved color-magnitude diagrams of nearby galaxies
where individual TP-AGB stars can be identified.  Fewer are observed than expected. 
\citet{Kriek10} fit \citet{BZ03} and \citet{maraston05} models to the SEDs of post-starburst galaxies. 
Both provide a good fit of the optical part of the spectrum ($\lambda < 6000\;\mathrm{\AA}$).  The \citet{BZ03} 
also fit the data at longer wavelengths, while the model of \citet{maraston05} over-predicts the luminosity in this
part of the spectrum (see their Figure~3).  Similarly, \citet{Zib13} sought to observe the strong spectral features
expected from TP-AGB stars in NIR spectra, but did not find them.  Modeling of the latest phases of stellar 
evolution does not yet appear to be sufficiently accurate to confidently predict the NIR spectra of
complex stellar populations.  

\begin{figure*}
\epsscale{1.0}
\plotone{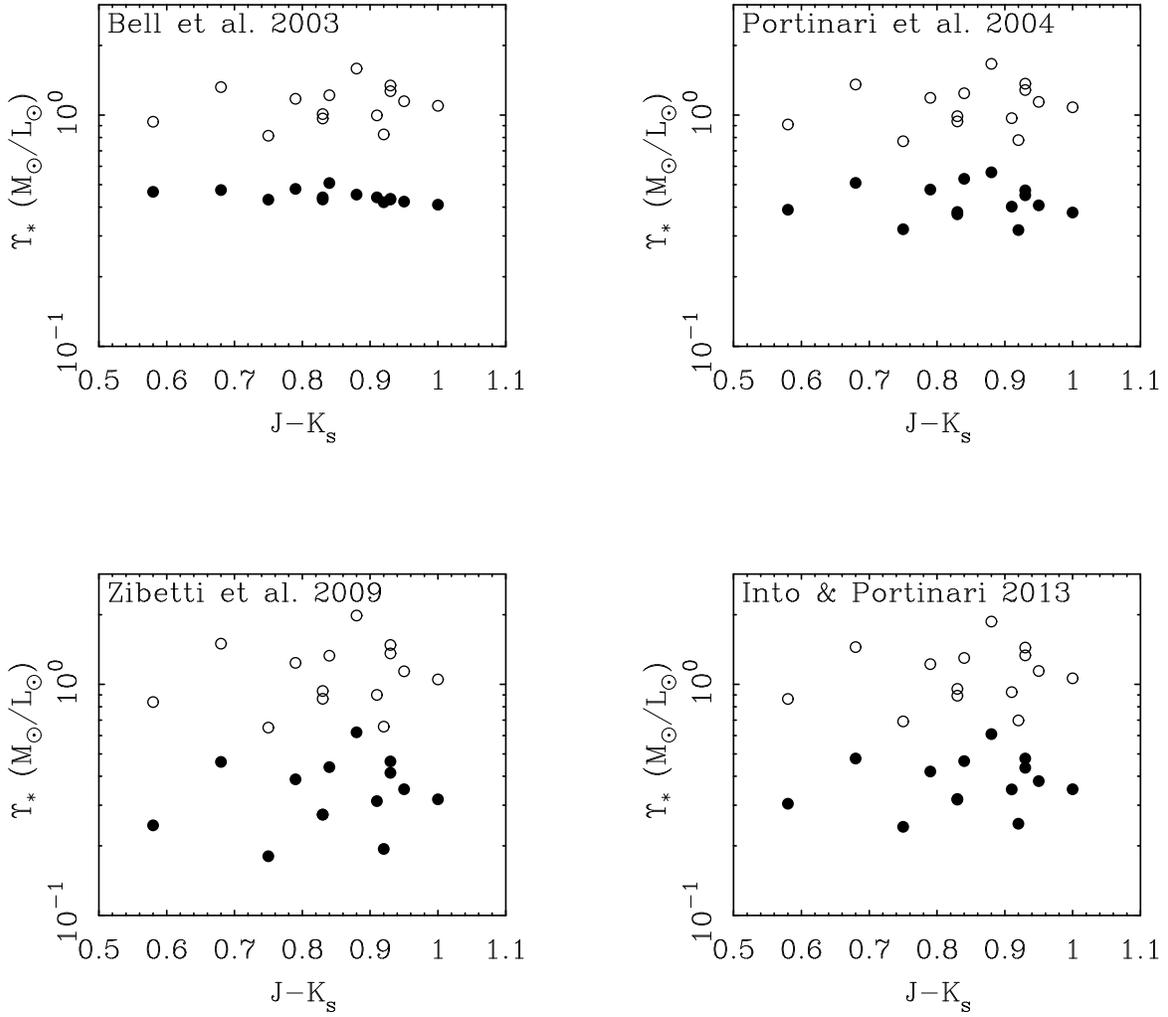}
\caption{Stellar mass-to-light ratios in the $I$-band (open circles) and [3.6] (filled circles) as a function of $J-K_s$ color.
The models of \citet{Bell03,Port04,Zib09,IP13} have been revised according to the fits in Table \ref{MtoMlines}.
Contrary to the case of $B-V$ (Fig.~\ref{MLpop_VIK}), there is little correlation:  $J-K_s$ does not provide a good indicator of \ML.
\label{MLpop_VIK_JK}}
\end{figure*}

In the mean time, considerable caution is warranted in assigning stellar masses based on SED fits \citep{conroy}.  
The results will depend not only on the model adopted, but also on the range of wavelengths fit.  
The offset between optical and NIR luminosity discussed here will result in a systematic skew towards
lower stellar masses as more NIR data are incorporated into SED fits.
Other fit parameters, like the star formation rate, will also be affected.

\section{Self Consistent CMLR}
\label{sec:CMLR}

Here we consider what is necessary to make empirically self-consistent CMLR.
We note first that all models provide a reasonably consistent picture in the optical.
This is not surprising given the optical heritage of the subject, and that the difficulty in modeling the latest stages of stellar evolution
mostly impacts the NIR.  We therefore adopt the $V$-band as a reference point that is well grounded in Galactic star counts \citep{flynn}.

\begin{deluxetable*}{lcccccccccc}
\tablewidth{0pt}
\tablecaption{Self-Consistent Population Synthesis CMLR}
\tablehead{
\colhead{Model} & \colhead{$a_V$} & \colhead{$b_V$}  & \colhead{$\alpha_{I}$}  & \colhead{$\beta_{I}$} & \colhead{$\alpha_{[3.6]}$} 
& \colhead{$\beta_{[3.6]}$} & \colhead{$\Upsilon_{0.6}^V$} & \colhead{$\Upsilon_{0.6}^I$} & \colhead{$\Upsilon_{0.6}^K$} & \colhead{$\Upsilon_{0.6}^{[3.6]}$}
}
\startdata
Bell \etal\ (2003)  & $-$0.628 & 1.305 & $-$0.259 & 0.565 & $-$0.313 & $-$0.043 & 1.43 & 1.20 & 0.60 & 0.46 \\
Portinari \etal\ (2004) &$-$0.654 & 1.290 & $-$0.302 & 0.644 & $-$0.575 & \phs0.394 & 1.32 & 1.22 & 0.60 & 0.46 \\
Zibetti \etal\ (2009)  & $-$1.075 & 1.837 & $-$0.446 & 0.915 & $-$1.115 & \phs1.172 & 1.07 & 1.27 & 0.50 & 0.39 \\
Into \& Portinari (2013) & $-$0.900 & 1.627 & $-$0.394 & 0.820 & $-$0.841 & \phs0.771 & 1.19 & 1.25 & 0.54 & 0.42
\enddata
\tablecomments{Stellar mass-to-light ratios in the $V$, $I$, and $K$-bands given by the formula $\log \ML^j = \alpha_j + \beta_j (B-V)$.  
For each model, the $V$-band is identical to that in Table \ref{popsynthML}, but the $I$ and [3.6] bands have been revised
to attain self-consistency with the $V$-band (see text).  The resulting lines are the fits shown in Fig.~\ref{MLpop_VIK}.
For reference, the mass-to-light ratio at $B-V = 0.6$ is also given. For $K_s$, we assume $K_s - [3.6] = 0.31$ (the mean observed
color) so that $\ML^K = 1.29 \ML^{[3.6]}$.}
\label{popsynthrevised} 
\end{deluxetable*}

\subsection{Self-Consistent Stellar Masses}
\label{sec:SCSM}

The relation between the mass computed in one band and that in another appears well defined, if not the desired 1:1 ratio (Fig.~\ref{MvM}).
We begin by fitting linear relations between the stellar mass in band $j = I$ and [3.6] and that in $V$ of the form
\begin{equation}
\log(\mathrm{M}_*^j/\mathrm{M}_0) = B_j \log(\mathrm{M}_*^V/\mathrm{M}_0).
\label{eq:MMtilt}
\end{equation}
Here $B_j$ is the slope of the fitted line, and M$_0$ the mass where the $V$-band and 
band $j$ intersect.  The fitted lines are shown in Fig.~\ref{MvM} and reported in Table \ref{MtoMlines}.

\begin{figure*}
\plotone{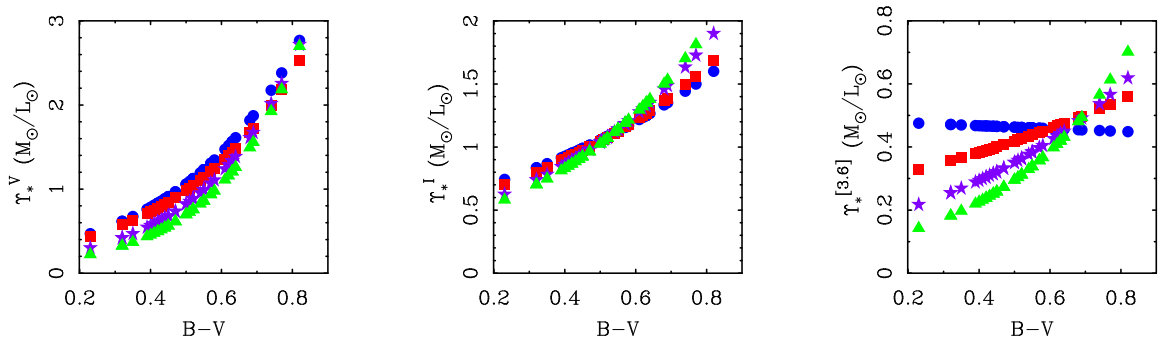}
\caption{The relation between $B-V$ color and the stellar mass-to-light ratio in the $V$-band (left), $I$-band (center) and the Spitzer [3.6] band (right) 
after correction of each stellar population model (Table \ref{popsynthrevised}) to obtain self-consistency.  Symbols as in Fig.\ \ref{MLcolor}.
The $V$-band panel is identical to that in Fig.\ \ref{MLcolor} as the models have been self-normalized to that band.
Agreement between the models is improved in the other bands, though perceptible differences persist.
\label{MLcolor_corrected}}
\end{figure*}

These lines provide a mapping between the mass in the reference $V$-band and that in the other filters.
They represent what is needed to obtain self-consistency within the context of each model.
They do not tell us what is right in an absolute sense, but they do tell us what would, on average,
return the same stellar mass from luminosities measured in each band.

Note that in all cases the slope $B_j > 1$.   
Higher mass galaxies are generally redder, so the slope $B$ presumably reflects a misestimate of the
color slopes $b_j$ tabulated in Table \ref{popsynthML}.  These are generally too large, in the sense that
the mass-to-light ratios in the redder bands of real galaxies do not vary as much with color as expected.

\subsection{Primary Color Dependence}
\label{sec:primary}

We use the lines fit in Fig.~\ref{MvM} and Table \ref{MtoMlines} to estimate a revised mass-to-light ratio for each galaxy in both $I$ and [3.6].
In effect, we assume that the mass indicated by the $V$-band is correct, and renormalize the other bands accordingly.
We then plot these against color to search for a revised CMLR that is self-consistent within the context of each model.  
The revised mass-to-light ratios are plotted
against $B-V$ in Fig.~\ref{MLpop_VIK}, $V-I$ in Fig.~\ref{MLpop_VIK_VI}, and $J-K_s$ in Fig.~\ref{MLpop_VIK_JK}.

There exist reasonably well defined CMLRs in Fig.~\ref{MLpop_VIK}:  \ML\ does correlate with $B-V$, if not quite with the expected slope.
The same cannot be said of the redder colors.  There is little if any perceptible slope of \ML\ with either $V-I$ (Fig.~\ref{MLpop_VIK_VI})
or $J-K_s$ (Fig.~\ref{MLpop_VIK_JK}), and a great deal of scatter in most cases.  While these colors may be useful as metallicity indicators \citep[e.g.,][]{BdJ00}, 
they appear to have little power to predict \ML.  This is consistent with the expectations of our own 
models \citep[\S \ref{sec:popsynth};][]{rakosschombert09a,SM2014b}, and with our unsatisfactory experience
in attempting to use these colors to build the equivalent of Table \ref{popsynth} (\S \ref{sec:galmass}).

\begin{figure*}
\epsscale{1.0}
\plotone{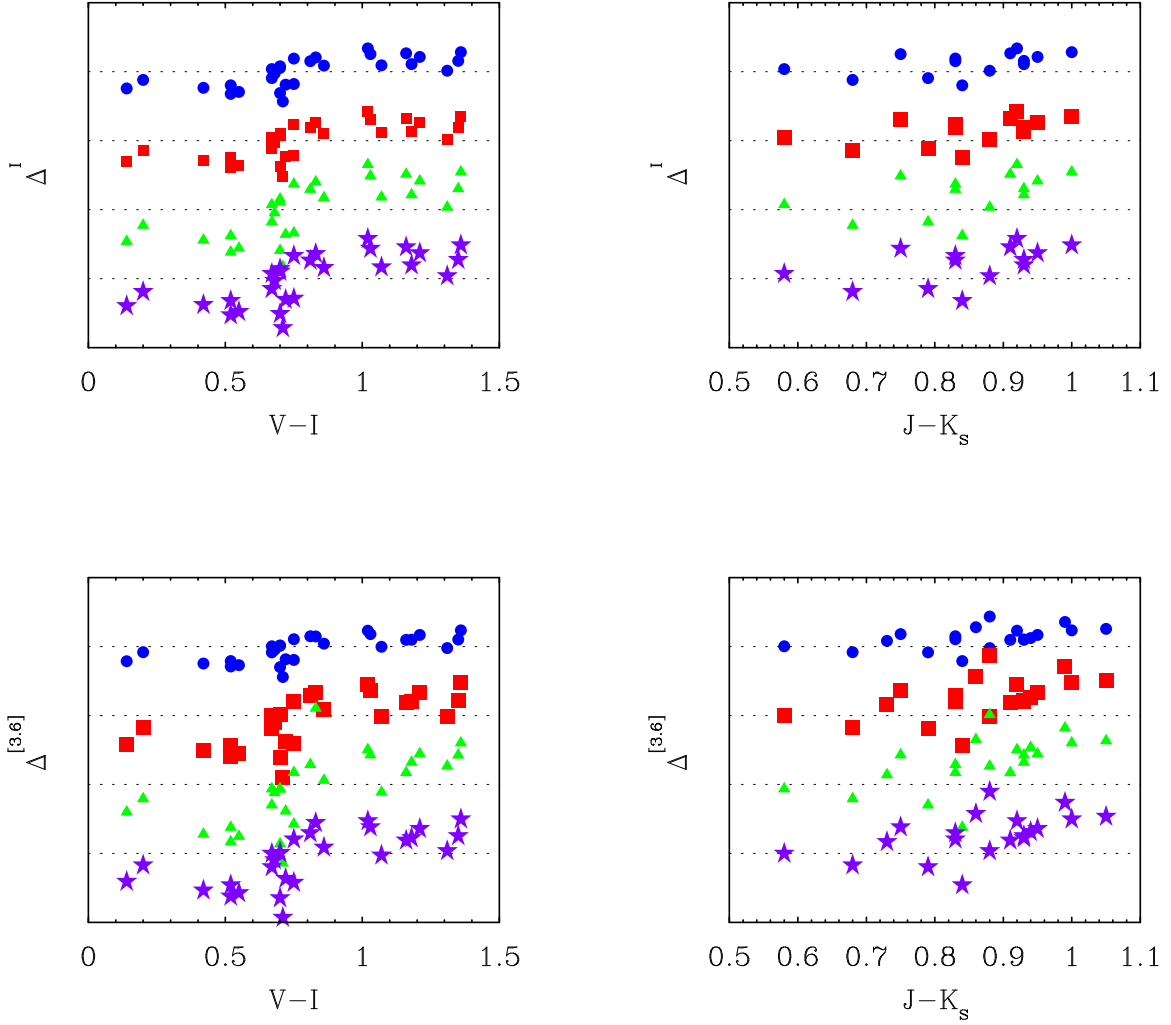}
\caption{The offset in mass-to-light ratios in the $I$-band (top row) and [3.6] (bottom row)
as a function of a second color: $V-I$ (left column) and $J-K$ (right column).
$\Delta$ is the logarithmic offset between the mass-to-light ratio that gives self-consistent stellar masses (Table \ref{MtoMlines})
and that approximated from the mean trend with $B-V$ color (Table \ref{popsynthrevised}).
The offsets for each population model have been shifted for clarity (symbols are per Fig.\ \ref{MLcolor}).
Zero offset is shown by the dotted lines, which are offset from each other by 0.2 dex.
Deviations of the data from these lines show when extra information about the
mass-to-light ratio is provided by the secondary color above and beyond that predicted by $B-V$ (Table \ref{secondcolorterm}).
The trend is as expected: redder colors indicate higher mass-to-light ratios.  The effect of $V-I$ is essentially binary: there is
an approximately constant shift in \ML\ above or below $V-I \approx 0.7$.  There is rather less information in $J-K_s$, though
very red galaxies do tend to have higher \ML.  The modest amplitude of these shifts imply that a single optical color 
contains most but not all of the information that can be used to constrain \ML.
}
\label{secondcolor}
\end{figure*}

We fit lines to the data in Fig.~\ref{MLpop_VIK} and provide the resulting self-consistent 
CMLR\footnote{The $B-V$ color maps closely to $g-r$ \citep{Jester2005}, so it should be straightforward to translate these CMLR into
SDSS bands if desired.} in Table~\ref{popsynthrevised}.
These are shown in Fig.~\ref{MLcolor_corrected}.  The agreement between models for the CMLR in the $I$-band is greatly improved.
That at [3.6] is also better, though considerable differences remain (compare Fig.~\ref{MLcolor_corrected} to Fig.~\ref{MLcolor}).

Among the four population synthesis models considered here, those of \citet{Bell03} and \citet{Port04} require the smallest corrections.
The corrections to the models of \citet{Zib09} and \citet{IP13} are rather larger.  
As anticipated, the revised slopes $\beta_j$ become shallower than the corresponding $b_j$ for all models in Table~\ref{popsynthML}.
The corrections to the older models are plausibly at the level one might expect.  For the newer models, both the slope and intercept of the
CMLR change substantially.  The revised CMLR in Fig.~\ref{MLpop_VIK} also has more scatter for the newer models than for the older
models.  It appears that the more recent models do not provide an improved description of real galaxies.

The most obvious culprit for the degraded performance of the newer models is an overestimate of the contribution
to the NIR light by TP-AGB stars.  These stars hardly affect the $V$-band while the $I$-band and [3.6] are both strongly
affected.  It appears that well-intentioned attempts to incorporate the latest evolutionary tracks for TP-AGB stars
have caused the more recent models to deviate further from reality than the preceding generation of models.  

Indeed, the results here can be used to inform future modeling efforts.
The self-consistent CMLR provide a benchmark for comparison to models, which should obtain the same spectral shape for a given stellar mass.
In addition to defining the general trend of \ML\ with color, the self-consistent CMLR might also help identify specific populations of stars
that may affect particular parts of the spectrum; TP-AGB stars being the obvious example here.

\subsection{Secondary Color Dependence}
\label{sec:secondary}

The $B-V$ color provides the best primary indicator of \ML\ among the available colors (Figs.~\ref{MLpop_VIK} -- \ref{MLpop_VIK_JK}).  
While we find $V-I$ and $J-K_s$
to be unsatisfactory in this regard, that does not mean they are completely devoid of information.  Here we search for improvements
to the primary CMLR by including these colors as a secondary term.

The CMLR in Table~\ref{popsynthrevised} is based on a fit against color of the stellar mass that is estimated through the procedure
described in \S \ref{sec:SCSM}.  As such, they are not guaranteed to perfectly reproduce the input.  Indeed, there
is a fair amount of scatter in Fig.~\ref{MLpop_VIK}, though some scatter is expected just from variations in the star formation history.

\begin{deluxetable*}{lcccccc}
\tablewidth{0pt}
\tablecaption{Second Color Correction Terms}
\tablehead{
\colhead{Model}
 & \multicolumn{2}{c}{$V-I < 0.65$} & \multicolumn{2}{c}{$V-I > 0.75$} & \multicolumn{2}{c}{$J-K > 0.90$} \\
 & \colhead{$\Delta^I$} & \colhead{$\Delta^{[3.6]}$} & \colhead{$\Delta^I$} & \colhead{$\Delta^{[3.6]}$}
  & \colhead{$\Delta^I$}  & \colhead{$\Delta^{[3.6]}$} 
}
\startdata
Bell \etal\ (2003)  & $-$0.047 & $-$0.044 & 0.036 & 0.024 & 0.046 & 0.037 \\
Portinari \etal\ (2004) & $-$0.057 & $-$0.089 & 0.045 & 0.048 & 0.057 & 0.075 \\
Zibetti \etal\ (2009)  & $-$0.089 & $-$0.118 & 0.070 & 0.075 & 0.089 & 0.099 \\
Into \& Portinari (2013) & $-$0.077 & $-$0.093 & 0.062 & 0.053 & 0.078 & 0.079
\enddata
\tablecomments{Corrections to the stellar mass-to-light ratio from a second color term, either $V-I$ or $J-K$ (not both).
The tabulated $\Delta$ for each band and color range can be added to the formula from Table \ref{popsynthrevised} 
to improve the estimate of stellar mass-to-light ratio: $\log \ML^j = \alpha_j + \beta_j (B-V) + \Delta^j$
when the second color is available.  $\Delta = 0$ for color ranges not listed:  $0.65 \le V-I \le 0.75$ and $J-K \le 0.9$.}
\label{secondcolorterm} 
\end{deluxetable*}

To check if some further improvement can be obtained, we define an offset $\Delta^j$ for $j = I$, [3.6]. This is simply the difference between the
mass-to-light ratio given by the relation in Table~\ref{MtoMlines} and that in Table~\ref{popsynthrevised}.  This is the residual between
the data and the line in Fig.~\ref{MLpop_VIK}.

\begin{deluxetable*}{lcccccccccc}
\tablewidth{0pt}
\tablecaption{Revised CMLR}
\tablehead{
\colhead{Model} & \colhead{$a_V$} & \colhead{$b_V$}  & \colhead{$\alpha_{I}$}  & \colhead{$\beta_{I}$} & \colhead{$\alpha_{[3.6]}$} 
& \colhead{$\beta_{[3.6]}$} & \colhead{$\Upsilon_{0.6}^V$} & \colhead{$\Upsilon_{0.6}^I$} & \colhead{$\Upsilon_{0.6}^K$} & \colhead{$\Upsilon_{0.6}^{[3.6]}$}
}
\startdata
Bell \etal\ (2003)  & $-$0.628 & 1.305 & $-$0.275 & 0.612 & $-$0.322 & $-$0.007 & 1.43 & 1.24 & 0.61 & 0.47 \\
Portinari \etal\ (2004) &$-$0.654 & 1.290 & $-$0.321 & 0.701 & $-$0.594 & \phs0.467 & 1.32 & 1.26 & 0.63 & 0.49 \\
Zibetti \etal\ (2009)  & $-$1.075 & 1.837 & $-$0.477 & 1.004 & $-$1.147 & \phs1.289 & 1.07 & 1.33 & 0.54 & 0.42  \\
Into \& Portinari (2013) & $-$0.900 & 1.627 & $-$0.421 & 0.898 & $-$0.861 & \phs0.849 & 1.19 & 1.31 & 0.58 & 0.45 
\enddata
\tablecomments{Stellar mass-to-light ratios in the $V$, $I$, and $K$-bands given by the formula $\log \ML^j = \alpha_j + \beta_j (B-V)$.  
For each model, the $V$-band is identical to that in Table \ref{popsynthML}, but the $I$ and [3.6] bands have been revised
to attain self-consistency with the $V$-band, and further corrected for $V-I$ as a second color term: the $\Delta$ of Table \ref{secondcolorterm}
have been incorporated to produce these revised CMLR.  
The resulting lines are fits to the data in Fig.~\ref{MLpoptweak_VIK}, providing our best estimate of the CMLR.
For reference, the mass-to-light ratio at $B-V = 0.6$ is also given.  For the $K_s$-band, 
we assume $\ML^K = 1.29 \ML^{[3.6]}$.  
}
\label{revisedCMLR} 
\end{deluxetable*}

We plot the residual offset $\Delta^j$ against $V-I$ and $J-K_s$ in Fig.~\ref{secondcolor}.
There is a clear effect with $V-I$: blue galaxies are offset to lower \ML, and red ones to higher \ML\ than nominally anticipated
by the CMLR of Table~\ref{popsynthrevised}.  There is little effect in $J-K_s$, though the reddest galaxies do show some offset.
In both cases, the effect goes in the expected\footnote{There is a hint in Fig.~\ref{MLpop_VIK_VI} that in the model of \citet{Bell03}
$\ML^{[3.6]}$ declines a small amount as $V-I$ becomes redder.  This is not apparent in the other models, and may be an artifact of
the slight overestimate of $\ML^{[3.6]}$ in this case (Fig.~\ref{MvM}).  This trend, if real, only happens when
$V-I$ is used as a primary indicator.  When used as a secondary indicator, galaxies that are redder at a given $B-V$ have higher \ML,
and bluer galaxies have lower \ML.} sense:  galaxies that are bluer in $V-I$ at a given $B-V$ have a lower \ML, and
those that are redder in either $V-I$ or $J-K_s$ at fixed $B-V$ have higher \ML.

\begin{figure*}
\epsscale{1.0}
\plotone{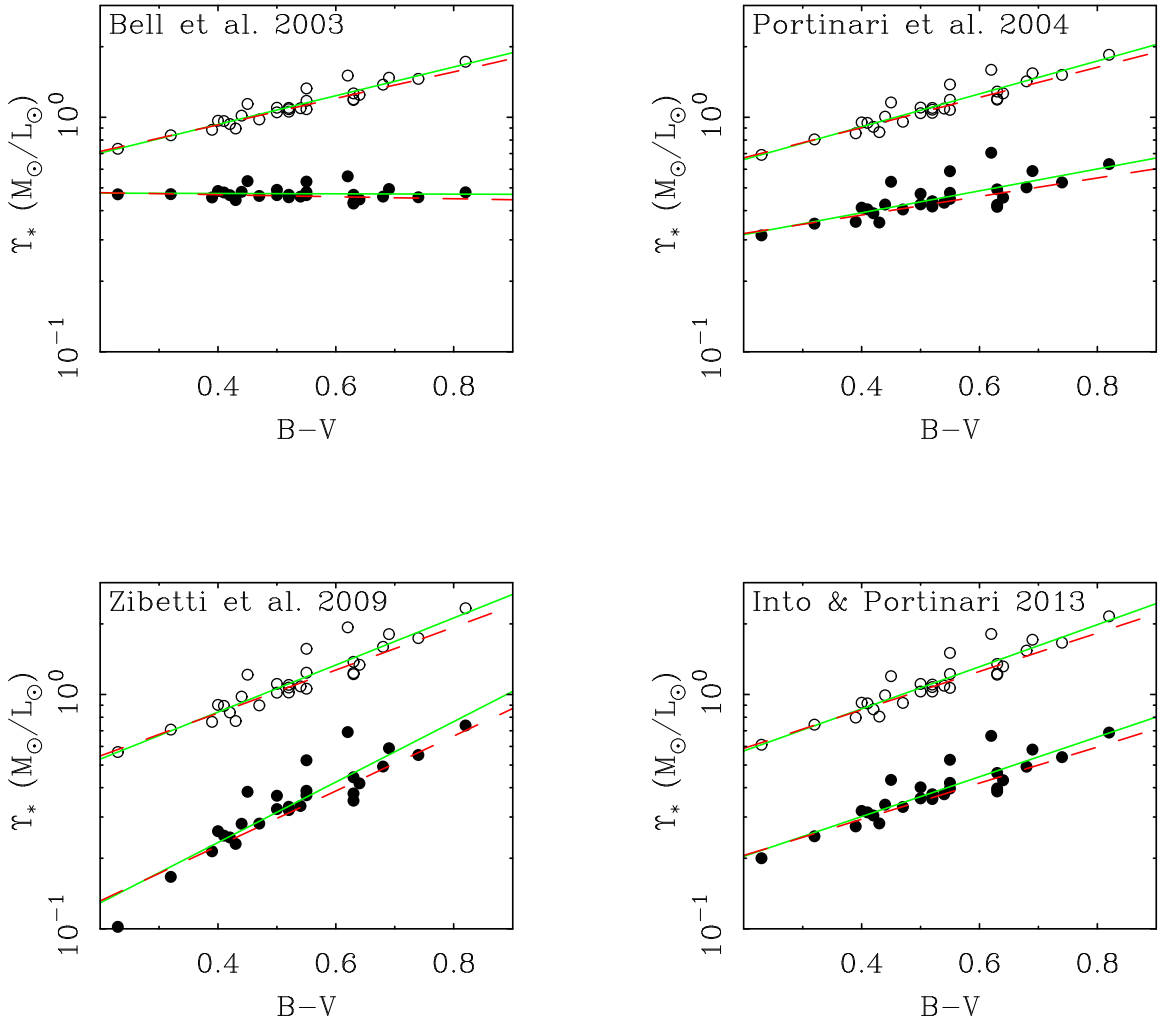}
\caption{Stellar mass-to-light ratios in the $I$-band (open circles) and [3.6] (filled circles) as in Fig.\ \ref{MLpop_VIK}
corrected with $V-I$ as a second color term as per Table \ref{secondcolorterm}.  Though the typical correction is less than 0.1 dex,
the reduction in scatter is noticeable. The data are re-fit (solid lines) to provide an improved CMLR (Table \ref{revisedCMLR}).
The previous fits from Fig.\ \ref{MLpop_VIK} (dashed lines) are reproduced as here for comparison.  
\label{MLpoptweak_VIK}}
\end{figure*}

We give a correction term as a function of color in Table \ref{secondcolorterm}.  We treat this term as a simple step function offset.
This describes the $V-I$ data quite well outside a narrow transition region.  The $J-K_s$ data give the visual impression of a linear
rise in $\Delta$ redwards of a long region of no effect, but the scatter is large enough that there is no perceptible improvement with
such a more complicated fit.  Indeed, we see little added value in $J-K_s$ as a \ML\ estimator.

There is clear value added in combining $V-I$ with $B-V$ as an indicator of \ML.
Fig.~\ref{MLpoptweak_VIK} shows the results of correcting the formulae in Table~\ref{popsynthrevised} with $\Delta$ from Table~\ref{secondcolorterm}.
The best fit line is only slightly changed; these slight revisions to the CMLR are given in Table.~\ref{revisedCMLR}.  
Perhaps the most remarkable change is that the scatter is greatly reduced, perhaps as much as one could reasonably hope.
Consequently, $B-V$ and $V-I$ appear to contain nearly all the information about \ML\ that the SED has to offer, 
at least for $\lambda > 4000\;\textrm{\AA}$ \citep[UV colors appears promising for elliptical galaxies:][]{Z14}.

\section{Discussion}
\label{sec:discuss}

\subsection{Estimating Stellar Mass-to-Light Ratios}
\label{sec:MLest}

Examining Fig.~\ref{MLpoptweak_VIK} and Table~\ref{revisedCMLR}, it becomes apparent that stellar masses consistent 
with the $V$-band estimates are only obtained if the mass-to-light ratios in the redder bands are relatively heavy.  
The absolute range in both $\ML^I$ and $\ML^{[3.6]}$ is now much narrower than originally predicted.
Comparing their value at a fiducial color of $B-V = 0.6$ in Tables~\ref{popsynthML} and \ref{revisedCMLR}, 
we find the range of the $I$-band mass-to-light ratio has changed from $0.76 < \ML^I < 1.25\;\MLsun$ to $1.24 < \ML^I < 1.33\;\MLsun$, 
while that at [3.6] has narrowed from $0.14 < \ML^{[3.6]} < 0.62\;\MLsun$ to $0.42 < \ML^{[3.6]} < 0.49\;\MLsun$.
Factors of two (or more) variation have been reduced to $< 20\%$.
This would appear to validate the long standing intuition that redder bands would provide the more direct measure of stellar mass.

Looking at the individual models, the revised CMLR based on the model of \citet{Bell03} has the least scatter.
The model of \citet{Port04} is very nearly as good. The scatter becomes progressively worse in the models of \citet{IP13}
and \citet{Zib09}.  This is not surprising since the revised CMLR of the latter have had gross corrections to both their intercept and slope.
These models did not have their NIR luminosities in the right ballpark to begin with.

Comparing the models of \citet{Bell03} and \citet{Port04}, the revised CMLR in the $I$-band are practically indistinguishable.
At [3.6], the two differ slightly in that the revised CMLR of the \citet{Bell03} model has effectively zero color dependence,
while that of \citet{Port04} does show a shallow slope.  The latter is non-zero at $\sim 2.5 \sigma$ significance, so there is
formally some slight tension between the models.  This seems a reasonable amount of agreement given the various uncertainties.
For example, we tried varying the prescription for internal extinction from zero to the RC3 \citep{RC3} prescription, with 
results intermediate between those in Tables \ref{popsynthrevised} and \ref{revisedCMLR}. These bracketing cases
are not much different, and such differences are much smaller than the uncertainty in the IMF.

As a practical matter, the mass-to-light ratio in the NIR is effectively constant.  The revised \citet{Bell03} model has 
$\langle \ML^{[3.6]} \rangle = 0.47\;\MLsun$.  All galaxies are within 0.1 dex of this mean value.  Comparing this with the model 
of \citet{Port04}, 22 of 28 galaxies are within 0.1 dex of $0.47\;\MLsun$, and none deviate by as much as 0.2 dex.
Even the model of \citet{Zib09}, with its different IMF, larger scatter, and much stronger color dependence is largely consistent with 
this mean value but for galaxies with extreme colors.

It therefore seems advisable to adopt $\ML^{[3.6]} = 0.47\;\MLsun$ as a characteristic value, with corresponding value $\ML^K = 0.6\;\MLsun$.  
There is no clear need to correct these mean values with color terms.
Indeed, the range of variation implied by the slope of the \citet{Port04} model is comparable to the scatter.
It therefore appears unwise to attempt any color correction in these bands, as one is as likely to add noise as to improve the situation.
A corollary is that a NIR image is already as good a map of the stellar mass as it is possible\footnote{We are aware of a variety of attempts
to build stellar mass maps by making color corrections to an image.  Indeed, we have exerted no small effort along these lines ourselves.
As well motivated as these attempts are, they do not appear capable of providing an improvement over the direct NIR image given 
the present state of model development.} to obtain.  

In principle, one would expect that by using all of the spectral information by fitting the complete SED, one would obtain the best stellar mass estimate.
In practice, this does not appear to be true: population synthesis models are not yet up to this task.  Mismatches between models and reality will inevitably
introduce systematic errors into any such procedure.  In practice, one is better off simply assuming a constant \ML\ in the NIR.

In contrast to the NIR, a clear color dependence persists in the I-band.  Unlike [3.6], the $I$-band luminosity does not provide a direct measure of stellar mass.
Nevertheless, Fig.~\ref{MLpoptweak_VIK} gives reason to hope that a simple color correction, as provided in Table \ref{revisedCMLR}, or the combination of 
Tables \ref{popsynthrevised} and \ref{secondcolorterm}, can be utilized to provide a reasonable estimate of stellar mass.  
This is not as good as measuring the NIR luminosity directly, but is often more readily obtained.

Note that a $B-V$ color is needed to estimate $\ML^I$; $V-I$ by itself is no help.  If only $B$ and $V$ are available without the $I$-band, then
one is back to relying on the population synthesis models, which appear to be fairly robust in the optical.  If only a single bandpass is available,
to a crude first approximation, $\ML \approx 1.2\;\MLsun$ in each of $B$, $V$, and $I$.

The absolute values of the mass-to-light ratios given here explicitly assume that the modeling of the optical portion of the spectrum is essentially
correct.  Really the data only constrain the ratio of optical-to-NIR luminosity to be higher than most models indicate so that
$\ML^V/\ML^{[3.6]} \approx 2.5$ -- 3.  One is free to adjust the normalization of both optical and NIR mass-to-light ratios so long as this 
ratio is preserved, for example by altering the IMF.

\subsection{Comparison with Independent Constraints}
\label{sec:IndCon}

There are independent constraints that we can check \ML\ against.  One obvious example is star counts in the Milky Way.
For the solar cylinder, \citet{flynn} measure $B-V = 0.58$ and $V-I = 0.90$ (their Table 5).  They estimate the stellar mass-to-light ratios
of the Milky Way to be $\ML^V = 1.5$ and $\ML^I = 1.2\;\MLsun$ (both $\pm 0.2$).  This compares favorably with the $B-V = 0.6$ values
in Table \ref{revisedCMLR}, where $\ML^V = 1.43$ and $\ML^I = 1.24\;\MLsun$ for the revised \citet{Bell03} CMLR.  Staying with this model
to be specific, we can apply the formulas we have derived to the observed colors.  Applying the formula from Table \ref{popsynthrevised}, 
which depends only on $B-V$, we obtain $\ML^I = 1.17\;\MLsun$.  Correcting this with $V-I$ as a second term as per Table \ref{secondcolorterm},
we obtain $\ML^I = 1.27\;\MLsun$.  The formula in Table \ref{revisedCMLR} (which depends only on $B-V$) gives $\ML^I = 1.20\;\MLsun$.
This gives an idea of the consistency and precision that can be obtained.

\citet{Eskew} calibrate the conversion between NIR flux and stellar mass in the LMC.  They obtain $\ML^{[3.6]} = 0.5\;\MLsun$,
in excellent agreement with our results.  \citet{Eskew} further discuss a $[3.6]-[4.5]$ color correction to this basic result.
Given the lack of sensitivity of \ML\ to $J-K_s$ that we find, and the small scatter (0.11 mag.) that we measure in $K_s - [3.6]$,
we find it unlikely that the $[3.6]-[4.5]$ color provides much information to improve estimations of \ML\ in the NIR.  Irrespective
of this detail, our basic results are in very good agreement.

Star count constraints, though direct, are still subject to uncertainty in the IMF.
We can also compare our results to dynamical constraints.  For example, \citet{bovyrix} have recently measured the vertical force in the
Milky Way disk over a substantial range of radii outside the solar circle.  Their results are consistent with the Milky Way model constructed
by \citet{M08} based on the work of \citet{flynn}.  Consequently, the implied mass-to-light ratios are also consistent.

The vertical force in external disk galaxies also provides a constraint.  Our results are simultaneously consistent and in conflict with those
of the disk mass survey \citep{bershady10}.  \citet{diskmass7} find that essentially all disk galaxies have indistinguishable mass-to-light ratios
in the $K$-band, just as we do.  However, their normalization is different: \citet{diskmass7} find $\langle \ML^K \rangle = 0.31 \pm 0.07$.
This is basically a factor of two lower than the corresponding values in Table \ref{revisedCMLR}.

The lower mass scale favored by the disk mass survey \citep[e.g.,][]{bershady11} could readily be obtained through a simple renormalization.
As mentioned previously, all that we require here is the correct ratio between optical and NIR luminosities.  Whether this can be reconciled with 
measurements on the IMF and other known dynamical constraints \citep[e.g.,][]{selwood1999} is beyond the scope of this work.  We see no
clear cut reason to prefer one mass scale over another, so a systematic uncertainty of a factor of $\sim 2$ persists in the absolute stellar mass scale. 

The Baryonic Tully-Fisher relation \citep{btforig} provides another constraint.  Using the calibration based on gas rich galaxies of \citet{M12},
we have checked the implied NIR mass-to-light ratios for those galaxies discussed here for which adequate kinematic data exist.  
This is an independent check, as the baryonic mass is estimated from the observed rotation velocity as calibrated by galaxies where the stars
do not contribute substantial systematic uncertainty to the baryonic mass budget \citep{M11}.  The resulting stellar mass estimate
follows from the kinematically estimated baryonic mass less the observed gas mass; it is in not informed by the purely photometric results here.  
We find typical NIR mass-to-light ratios in the range $\ML^{[3.6]} = 0.4$ -- $0.5\;\MLsun$, consistent with our results in Table \ref{revisedCMLR}
(and by implication, heavier than those of the disk mass survey, but consistent with the Milky Way and LMC).  
We will explore this further in a companion paper.

\section{Conclusions}
\label{sec:conc}

We have used Spitzer data for disk galaxies spanning ten magnitudes in [3.6] absolute magnitude
to test the self-consistency of stellar population synthesis models from the optical to NIR bands.
Our main conclusions can be summarized as follows:
\begin{itemize}
\item Many commonly utilized stellar population models are not self-consistent in the sense that application of the same model to the same
galaxy results in different stellar masses depending on whether an optical or NIR luminosity is used.  
\item Ab initio models tend to overestimate the NIR luminosity relative to the optical luminosity for a given mass of stars.
\item Models adopting recent prescriptions for TP-AGB stars severely overstate the NIR luminosity.  
\item Self-consistency between optical and NIR observations can be achieved if NIR mass-to-light ratios are approximately constant.  
\item The typical value for self-consistency is $0.47\;\MLsun$ at $3.6\micron$ (equivalent to $0.6\;\MLsun$ in the $K_s$-band).  
\item The mass-to-light ratio in optical bands does depend on $B-V$ color (see Table \ref{revisedCMLR}).  Redder colors like $J-K_s$ carry little additional information.
\end{itemize}

Workers wishing to estimate the stellar masses of galaxies would do well to adopt a constant NIR mass-to-light ratio as calibrated here.
If NIR bands such as $K_s$ or [3.6] are not available, the mass-to-light ratio in bluer bands like $I$ do correlate with a $B-V$ color.
This color contains most of the information about \ML.  A slight improvement can be gained by using $V-I$ as a secondary indicator.
Redder colors like this and $J-K_s$ have themselves no power as primary predictors of \ML.

Once a NIR luminosity is measured, there appears to be little added value in fitting the complete SED so far as constraining the stellar mass goes.  
Indeed, such fits can only be as good as the population model the data are fit to.  Given the systematic offsets in the models found here, 
such SED fits are bound to suffer from systematic errors that will depend on the specific model employed and also on the range of wavelengths fit.
On the other hand, the results found here can be used to inform improvements in the models.

\acknowledgements  
We thank Laura Portinari and the referee for numerous helpful suggestions.
This work is based in part on observations made with the Spitzer Space Telescope, which is operated by the Jet Propulsion Laboratory, 
California Institute of Technology under a contract with NASA. Support for this work was provided by NASA through an award issued by 
JPL/Caltech. Other aspects of this work were supported in part by NASA ADAP grant NNX11AF89G and NSF grant AST 0908370.  
This research has made use of the NASA/IPAC Extragalactic Database (NED) which is operated by the Jet Propulsion Laboratory, 
California Institute of Technology, under contract with the National Aeronautics and Space Administration.

\bibliography{stms_new}

\begin{thebibliography}{53}
\expandafter\ifx\csname natexlab\endcsname\relax\def\natexlab#1{#1}\fi

\bibitem[{{Bell} \& {de Jong}(2000)}]{BdJ00}
{Bell}, E.~F., \& {de Jong}, R.~S. 2000, \mnras, 312, 497

\bibitem[{{Bell} \& {de Jong}(2001)}]{BdJ01}
---. 2001, \apj, 550, 212

\bibitem[{{Bell} {et~al.}(2003){Bell}, {McIntosh}, {Katz}, \&
  {Weinberg}}]{Bell03}
{Bell}, E.~F., {McIntosh}, D.~H., {Katz}, N., \& {Weinberg}, M.~D. 2003, \apjs,
  149, 289

\bibitem[{{Bershady} {et~al.}(2011){Bershady}, {Martinsson}, {Verheijen},
  {Westfall}, {Andersen}, \& {Swaters}}]{bershady11}
{Bershady}, M.~A., {Martinsson}, T.~P.~K., {Verheijen}, M.~A.~W., {Westfall},
  K.~B., {Andersen}, D.~R., \& {Swaters}, R.~A. 2011, \apjl, 739, L47

\bibitem[{{Bershady} {et~al.}(2010){Bershady}, {Verheijen}, {Swaters},
  {Andersen}, {Westfall}, \& {Martinsson}}]{bershady10}
{Bershady}, M.~A., {Verheijen}, M.~A.~W., {Swaters}, R.~A., {Andersen}, D.~R.,
  {Westfall}, K.~B., \& {Martinsson}, T. 2010, \apj, 716, 198

\bibitem[{{Bovy} \& {Rix}(2013)}]{bovyrix}
{Bovy}, J., \& {Rix}, H.-W. 2013, \apj, 779, 115

\bibitem[{{Bruzual} \& {Charlot}(2003)}]{BZ03}
{Bruzual}, G., \& {Charlot}, S. 2003, \mnras, 344, 1000

\bibitem[{{Chabrier}(2003)}]{chabrier}
{Chabrier}, G. 2003, \apjl, 586, L133

\bibitem[{{Conroy} \& {Gunn}(2010)}]{conroy}
{Conroy}, C., \& {Gunn}, J.~E. 2010, \apj, 712, 833

\bibitem[{{Dale} {et~al.}(2005){Dale}, {Bendo}, {Engelbracht}, {Gordon},
  {Regan}, {Armus}, {Cannon}, {Calzetti}, {Draine}, {Helou}, {Joseph},
  {Kennicutt}, {Li}, {Murphy}, {Roussel}, {Walter}, {Hanson}, {Hollenbach},
  {Jarrett}, {Kewley}, {Lamanna}, {Leitherer}, {Meyer}, {Rieke}, {Rieke},
  {Sheth}, {Smith}, \& {Thornley}}]{DSINGS}
{Dale}, D.~A., {et~al.} 2005, \apj, 633, 857

\bibitem[{{Dale} {et~al.}(2007){Dale}, {Gil de Paz}, {Gordon}, {Hanson},
  {Armus}, {Bendo}, {Bianchi}, {Block}, {Boissier}, {Boselli}, {Buckalew},
  {Buat}, {Burgarella}, {Calzetti}, {Cannon}, {Engelbracht}, {Helou},
  {Hollenbach}, {Jarrett}, {Kennicutt}, {Leitherer}, {Li}, {Madore}, {Martin},
  {Meyer}, {Murphy}, {Regan}, {Roussel}, {Smith}, {Sosey}, {Thilker}, \&
  {Walter}}]{Dale07}
---. 2007, \apj, 655, 863

\bibitem[{{de Blok} {et~al.}(1995){de Blok}, {van der Hulst}, \&
  {Bothun}}]{dBHB95}
{de Blok}, W.~J.~G., {van der Hulst}, J.~M., \& {Bothun}, G.~D. 1995, \mnras,
  274, 235

\bibitem[{{de Blok} {et~al.}(2008){de Blok}, {Walter}, {Brinks},
  {Trachternach}, {Oh}, \& {Kennicutt}}]{THINGS}
{de Blok}, W.~J.~G., {Walter}, F., {Brinks}, E., {Trachternach}, C., {Oh},
  S.-H., \& {Kennicutt}, R.~C. 2008, \aj, 136, 2648

\bibitem[{{de Vaucouleurs} {et~al.}(1991){de Vaucouleurs}, {de Vaucouleurs},
  {Corwin}, {Buta}, {Paturel}, \& {Fouqu{\'e}}}]{RC3}
{de Vaucouleurs}, G., {de Vaucouleurs}, A., {Corwin}, Jr., H.~G., {Buta},
  R.~J., {Paturel}, G., \& {Fouqu{\'e}}, P. 1991, {Third Reference Catalogue of
  Bright Galaxies.} (Springer, New York, NY)

\bibitem[{{Eskew} {et~al.}(2012){Eskew}, {Zaritsky}, \& {Meidt}}]{Eskew}
{Eskew}, M., {Zaritsky}, D., \& {Meidt}, S. 2012, \aj, 143, 139

\bibitem[{{Fisher} \& {Drory}(2008)}]{FD08}
{Fisher}, D.~B., \& {Drory}, N. 2008, \aj, 136, 773

\bibitem[{{Flynn} {et~al.}(2006){Flynn}, {Holmberg}, {Portinari}, {Fuchs}, \&
  {Jahrei{\ss}}}]{flynn}
{Flynn}, C., {Holmberg}, J., {Portinari}, L., {Fuchs}, B., \& {Jahrei{\ss}}, H.
  2006, \mnras, 372, 1149

\bibitem[{{Into} \& {Portinari}(2013)}]{IP13}
{Into}, T., \& {Portinari}, L. 2013, \mnras, 430, 2715

\bibitem[{{Jarrett} {et~al.}(2000){Jarrett}, {Chester}, {Cutri}, {Schneider},
  {Skrutskie}, \& {Huchra}}]{2MASS}
{Jarrett}, T.~H., {Chester}, T., {Cutri}, R., {Schneider}, S., {Skrutskie}, M.,
  \& {Huchra}, J.~P. 2000, \aj, 119, 2498

\bibitem[{{Jester} {et~al.}(2005){Jester}, {Schneider}, {Richards}, {Green},
  {Schmidt}, {Hall}, {Strauss}, {Vanden Berk}, {Stoughton}, {Gunn},
  {Brinkmann}, {Kent}, {Smith}, {Tucker}, \& {Yanny}}]{Jester2005}
{Jester}, S., {et~al.} 2005, \aj, 130, 873

\bibitem[{{Kennicutt} {et~al.}(2003){Kennicutt}, {Armus}, {Bendo}, {Calzetti},
  {Dale}, {Draine}, {Engelbracht}, {Gordon}, {Grauer}, {Helou}, {Hollenbach},
  {Jarrett}, {Kewley}, {Leitherer}, {Li}, {Malhotra}, {Regan}, {Rieke},
  {Rieke}, {Roussel}, {Smith}, {Thornley}, \& {Walter}}]{KSINGS}
{Kennicutt}, Jr., R.~C., {et~al.} 2003, \pasp, 115, 928

\bibitem[{{Kim} {et~al.}(2012){Kim}, {Im}, {Lee}, {Lee}, {Jun}, {Nakagawa},
  {Matsuhara}, {Wada}, {Oyabu}, {Takagi}, {Inami}, {Ohyama}, {Yamada}, {Helou},
  {Armus}, \& {Shi}}]{JHKim2012}
{Kim}, J.~H., {et~al.} 2012, \apj, 760, 120

\bibitem[{{Kriek} {et~al.}(2010){Kriek}, {Labb{\'e}}, {Conroy}, {Whitaker},
  {van Dokkum}, {Brammer}, {Franx}, {Illingworth}, {Marchesini}, {Muzzin},
  {Quadri}, \& {Rudnick}}]{Kriek10}
{Kriek}, M., {et~al.} 2010, \apjl, 722, L64

\bibitem[{{Kroupa}(1998)}]{kroupa}
{Kroupa}, P. 1998, in Astronomical Society of the Pacific Conference Series,
  Vol. 134, Brown Dwarfs and Extrasolar Planets, ed. R.~{Rebolo}, E.~L.
  {Martin}, \& M.~R. {Zapatero Osorio}, 483

\bibitem[{{Le Borgne} {et~al.}(2004){Le Borgne}, {Rocca-Volmerange},
  {Prugniel}, {Lan{\c c}on}, {Fioc}, \& {Soubiran}}]{PEGASE}
{Le Borgne}, D., {Rocca-Volmerange}, B., {Prugniel}, P., {Lan{\c c}on}, A.,
  {Fioc}, M., \& {Soubiran}, C. 2004, \aap, 425, 881

\bibitem[{{Makarova}(1999)}]{Mak99}
{Makarova}, L. 1999, \aaps, 139, 491

\bibitem[{{Maraston}(2005)}]{maraston05}
{Maraston}, C. 2005, \mnras, 362, 799

\bibitem[{{Marigo} {et~al.}(2008){Marigo}, {Girardi}, {Bressan}, {Groenewegen},
  {Silva}, \& {Granato}}]{Marigo2008}
{Marigo}, P., {Girardi}, L., {Bressan}, A., {Groenewegen}, M.~A.~T., {Silva},
  L., \& {Granato}, G.~L. 2008, \aap, 482, 883

\bibitem[{{Martinsson} {et~al.}(2013){Martinsson}, {Verheijen}, {Westfall},
  {Bershady}, {Andersen}, \& {Swaters}}]{diskmass7}
{Martinsson}, T.~P.~K., {Verheijen}, M.~A.~W., {Westfall}, K.~B., {Bershady},
  M.~A., {Andersen}, D.~R., \& {Swaters}, R.~A. 2013, \aap, 557, A131

\bibitem[{{McGaugh}(2008)}]{M08}
{McGaugh}, S.~S. 2008, \apj, 683, 137

\bibitem[{{McGaugh}(2011)}]{M11}
---. 2011, Physical Review Letters, 106, 121303

\bibitem[{{McGaugh}(2012)}]{M12}
---. 2012, \aj, 143, 40

\bibitem[{{McGaugh} \& {Bothun}(1994)}]{MB94}
{McGaugh}, S.~S., \& {Bothun}, G.~D. 1994, \aj, 107, 530

\bibitem[{{McGaugh} {et~al.}(2000){McGaugh}, {Schombert}, {Bothun}, \& {de
  Blok}}]{btforig}
{McGaugh}, S.~S., {Schombert}, J.~M., {Bothun}, G.~D., \& {de Blok}, W.~J.~G.
  2000, \apjl, 533, L99

\bibitem[{{Meidt} {et~al.}(2012){Meidt}, {Schinnerer}, {Knapen}, {Bosma},
  {Athanassoula}, {Sheth}, {Buta}, {Zaritsky}, {Laurikainen}, {Elmegreen},
  {Elmegreen}, {Gadotti}, {Salo}, {Regan}, {Ho}, {Madore}, {Hinz}, {Skibba},
  {Gil de Paz}, {Mu{\~n}oz-Mateos}, {Men{\'e}ndez-Delmestre}, {Seibert}, {Kim},
  {Mizusawa}, {Laine}, \& {Comer{\'o}n}}]{Meidt2012}
{Meidt}, S.~E., {et~al.} 2012, \apj, 744, 17

\bibitem[{{Melbourne} {et~al.}(2012){Melbourne}, {Williams}, {Dalcanton},
  {Rosenfield}, {Girardi}, {Marigo}, {Weisz}, {Dolphin}, {Boyer}, {Olsen},
  {Skillman}, \& {Seth}}]{Melbourne12}
{Melbourne}, J., {et~al.} 2012, \apj, 748, 47

\bibitem[{{Mu{\~n}oz-Mateos} {et~al.}(2009){Mu{\~n}oz-Mateos}, {Gil de Paz},
  {Zamorano}, {Boissier}, {Dale}, {P{\'e}rez-Gonz{\'a}lez}, {Gallego},
  {Madore}, {Bendo}, {Boselli}, {Buat}, {Calzetti}, {Moustakas}, \&
  {Kennicutt}}]{MM09}
{Mu{\~n}oz-Mateos}, J.~C., {et~al.} 2009, \apj, 703, 1569

\bibitem[{{Oh} {et~al.}(2008){Oh}, {de Blok}, {Walter}, {Brinks}, \&
  {Kennicutt}}]{OHTHINGS}
{Oh}, S., {de Blok}, W.~J.~G., {Walter}, F., {Brinks}, E., \& {Kennicutt},
  R.~C. 2008, \aj, 136, 2761

\bibitem[{{Portinari} {et~al.}(2004){Portinari}, {Sommer-Larsen}, \&
  {Tantalo}}]{Port04}
{Portinari}, L., {Sommer-Larsen}, J., \& {Tantalo}, R. 2004, \mnras, 347, 691

\bibitem[{{Schlafly} \& {Finkbeiner}(2011)}]{SF2011}
{Schlafly}, E.~F., \& {Finkbeiner}, D.~P. 2011, \apj, 737, 103

\bibitem[{{Schombert}(2011)}]{ARCHANGEL}
{Schombert}, J. 2011, Astrophysics Source Code Library, 7011

\bibitem[{{Schombert} {et~al.}(2011){Schombert}, {Maciel}, \&
  {McGaugh}}]{SMM11}
{Schombert}, J., {Maciel}, T., \& {McGaugh}, S. 2011, Advances in Astronomy,
  2011

\bibitem[{{Schombert} \& {McGaugh}(2014{\natexlab{a}})}]{SM2014b}
{Schombert}, J., \& {McGaugh}, S. 2014{\natexlab{a}}, submitted

\bibitem[{{Schombert} \& {Rakos}(2009)}]{rakosschombert09a}
{Schombert}, J., \& {Rakos}, K. 2009, \aj, 137, 528

\bibitem[{{Schombert} \& {McGaugh}(2014{\natexlab{b}})}]{SM2014}
{Schombert}, J.~M., \& {McGaugh}, S. 2014{\natexlab{b}}, Pub. Astron. Soc.
  Australia, 31, 11

\bibitem[{{Sellwood}(1999)}]{selwood1999}
{Sellwood}, J.~A. 1999, in Astronomical Society of the Pacific Conference
  Series, Vol. 182, Galaxy Dynamics - A Rutgers Symposium, ed. D.~R. {Merritt},
  M.~{Valluri}, \& J.~A. {Sellwood}, 351

\bibitem[{{Trachternach} {et~al.}(2009){Trachternach}, {de Blok}, {McGaugh},
  {van der Hulst}, \& {Dettmar}}]{trach}
{Trachternach}, C., {de Blok}, W.~J.~G., {McGaugh}, S.~S., {van der Hulst},
  J.~M., \& {Dettmar}, R. 2009, \aap, 505, 577

\bibitem[{{Tully} {et~al.}(2009){Tully}, {Rizzi}, {Shaya}, {Courtois},
  {Makarov}, \& {Jacobs}}]{EDD}
{Tully}, R.~B., {Rizzi}, L., {Shaya}, E.~J., {Courtois}, H.~M., {Makarov},
  D.~I., \& {Jacobs}, B.~A. 2009, \aj, 138, 323

\bibitem[{{Verheijen}(2001)}]{verhTF}
{Verheijen}, M.~A.~W. 2001, \apj, 563, 694

\bibitem[{{Walter} {et~al.}(2008){Walter}, {Brinks}, {de Blok}, {Bigiel},
  {Kennicutt}, {Thornley}, \& {Leroy}}]{FTHINGS}
{Walter}, F., {Brinks}, E., {de Blok}, W.~J.~G., {Bigiel}, F., {Kennicutt},
  R.~C., {Thornley}, M.~D., \& {Leroy}, A. 2008, \aj, 136, 2563

\bibitem[{{Zaritsky} {et~al.}(2014){Zaritsky}, {Gil de Paz}, \&
  {Bouquin}}]{Z14}
{Zaritsky}, D., {Gil de Paz}, A., \& {Bouquin}, A.~Y.~K. 2014, \apjl, 780, L1

\bibitem[{{Zibetti} {et~al.}(2009){Zibetti}, {Charlot}, \& {Rix}}]{Zib09}
{Zibetti}, S., {Charlot}, S., \& {Rix}, H.-W. 2009, \mnras, 400, 1181

\bibitem[{{Zibetti} {et~al.}(2013){Zibetti}, {Gallazzi}, {Charlot}, {Pierini},
  \& {Pasquali}}]{Zib13}
{Zibetti}, S., {Gallazzi}, A., {Charlot}, S., {Pierini}, D., \& {Pasquali}, A.
  2013, \mnras, 428, 1479

\end{thebibliography}

\bibliographystyle{apj}

\end{document}